\documentclass[fleqn,usenatbib]{mnras}
\UseRawInputEncoding
\usepackage{xcolor}
\usepackage{newtxtext,newtxmath}
 
\usepackage[T1]{fontenc}

\DeclareRobustCommand{\VAN}[3]{#2}
\let\VANthebibliography\thebibliography
\def\thebibliography{\DeclareRobustCommand{\VAN}[3]{##3}\VANthebibliography}

\usepackage{graphicx}	
\usepackage{amssymb}
\usepackage{amsmath}	
\usepackage{threeparttable}

\definecolor{address}{rgb}{0.36,0.54,0.66}
\definecolor{red}{rgb}{0.8,0.,0.}

\newcommand{\HI}{\mathrm{H}\,\textsc{\large{i}}}
\newcommand{\HIs}{\mathrm{H}\,\textsc{\textmd{i}}}

\newcommand{\Hmolb}{\mathbf{H_2}}
\newcommand{\Hmol}{\mathrm{H_2}}

\newcommand{\cc}{\mathrm{cm}^{-3}}
\newcommand{\K}{\mathrm{K}}
\newcommand{\eagle}{\textsc{\large eagle}{ }}
\newcommand{\eagleNS}{\textsc{\large eagle}}
\graphicspath{{Figures/}}

\title[Impact of cluster environment on $\Hmol$ content]{The relationship between cluster environment and molecular gas content of star-forming galaxies in the \textsc{\Huge eagle} simulation}
\author[A. Manuwal and A. R. H. Stevens]{Aditya Manuwal$^{1}$\thanks{E-mail: aditya.manuwal@icrar.org} and
Adam R. H. Stevens$^{1}$
\\
$^{1}$International Centre for Radio Astronomy Research, The University of Western Australia, 35 Stirling Highway, Crawley, WA 6009, Australia\\
}
\date{Accepted 2023 May 22. Received 2023 April 27; in original form 2022 December 23}
\pubyear{2022}

\begin{document}
\label{firstpage}
\pagerange{\pageref{firstpage}--\pageref{lastpage}}
\maketitle

\begin{abstract}
We employ the \textsc{\Large eagle} hydrodynamical simulation to uncover the relationship between cluster environment and $\Hmol$ content of star-forming galaxies at redshifts spanning $0\leq z\leq 1$. To do so, we divide the star-forming sample into those that are bound to clusters and those that are not. We find that, at any given redshift, the galaxies in clusters generally have less $\Hmol$ than their non-cluster counterparts with the same stellar mass (corresponding to an offset of $\lesssim 0.5$~dex), but this offset varies with stellar mass and is virtually absent at $M_\star\lesssim10^{9.3}~{\rm M}_\odot$. The $\Hmol$ deficit in star-forming cluster galaxies can be traced back to a decline in their $\Hmol$ content that commenced after first infall into a cluster, which occurred later than a typical cluster galaxy. Evolution of the full cluster population after infall is generally consistent with `slow-then-rapid' quenching, but galaxies with $M_\star\lesssim 10^{9.5}~{\rm M}_\odot$ exhibit rapid quenching. Unlike most cluster galaxies, 
star-forming ones were not pre-processed in groups prior to being accreted by clusters. For both of these cluster samples, the star formation efficiency remained oblivious to the infall. We track the particles associated with star-forming cluster galaxies and attribute the drop in $\Hmol$ mass after infall to poor replenishment, depletion due to star formation, and stripping of $\Hmol$ in cluster environments. These results provide predictions for future surveys, along with support and theoretical insights for existing molecular gas observations that suggest there is less $\Hmol$ in cluster galaxies.
\end{abstract}




\section{Introduction}
Star formation proceeds in cold and dense molecular clouds \citep{McKee2007,Kennicutt2012}, where 
high densities are conducive for processes that protect the gas against UV photodissociation \citep{Glover2012}, such as dust shielding and self-shielding within molecular clouds. Observations show that 
the star formation surface density is well correlated with the molecular gas surface density within galaxies 
in the local Universe, both on global \citep{Kennicutt1998} and local scales \citep{Bigiel2008}. 
This implies that the evolution of galaxies is regulated by 
the efficiency with which they convert molecular gas into stars 
(known as the `star formation efficiency'). As a result, 
one of the primary objectives of extragalactic astrophysics has been to understand the factors that govern the star formation efficiency of galaxies. The two key parameters that are relevant in this regard 
are the star formation rate (SFR) and molecular gas content of galaxies. 

Since molecular hydrogen ($\Hmol$) is the most abundant molecule in the interstellar medium (ISM), understanding its evolution is crucial for developing a comprehensive picture of the formation and evolution of galaxies. $\Hmol$, however, is not a good observational tracer of molecular gas because it lacks a permanent dipole moment, which renders its dipolar rotational transitions forbidden and precludes observing it in emission. Fortunately, molecular gas clouds also contain carbon and oxygen, which combine to form the CO molecule under the conditions that are prevalent in molecular clouds \citep{Van1988}. This molecule possesses a weak permanent dipole moment that allows for its excitation to higher rotational states, causing transitions like the $J=1\rightarrow 0$ emission which -- owing to its high atmospheric transparency and intensity -- is often used as an indirect tracer of $\Hmol$.

In practice, the $\Hmol$ fraction of a cloud is gauged by multiplying the CO luminosity by a CO-to-$\Hmol$ conversion factor ($\alpha_{\rm CO}$), which depends on ionisation conditions in the cloud and its environment \citep{Feldmann2012,Narayanan2012,Clark2015,Gong2018,Hu2022}. Initial conversion factors adopted in the literature were either calibrated 
to the Milky Way or other galaxies in the local Universe \citep[see the review by][]{Bolatto2013}.
Some studies have even extended this to $z\approx3$ \citep[e.g.][]{Genzel2015}. 
However, the factors that govern $\alpha_{\rm CO}$ are difficult to accurately account for, and 
the derived $\Hmol$ content of observed galaxies can carry significant intrinsic uncertainty. 
This is an important hindrance in accurate estimation of the $\Hmol$ content of metal-poor systems, 
where much of the carbon is ionised due to significant photodissociation of CO molecules by far-ultraviolet photons \citep{Bolatto2013}. CO can also be destroyed through collisions with $\rm{He}\,\textsc{\large ii}$ present in cosmic rays \citep{Bisbas2015}. For systems at high redshifts ($z>0.5$), there is an additional uncertainty introduced due to the fact that only higher CO rotational transitions are currently observable, which have to be corrected for gas excitation states to obtain the equivalent $J=1\rightarrow 0$ flux. The resulting uncertainties can be reduced to some extent by using complementary observations of [$\rm{C}\,\textsc{\large i}$]\footnote{The square brackets denote spontaneous emission.}\citep[e.g.][]{Valentino2018,Heintz2020}, $\rm{C}\,\textsc{\large ii}$ \citep[e.g.][]{Zanella2018}, dust extinction in optical/UV regime \citep[e.g.][]{Concas2019,Yesuf2019,Piotrowska2020}, and dust emission \citep[e.g.][]{Scoville2016}; although the latter may not be suitable for high redshifts due to the decline in dust content with metallicity. 

It is well established that galaxies in dense environments, like clusters, have different evolutionary tracks compared to their non-cluster counterparts.
This occurs because cluster galaxies are subject to mechanisms such as the stripping of gas through ram pressure exerted by the intracluster medium \citep{Gunn1972,Quilis2000}, thermal evaporation of gas due to interaction with hot intergalactic gas \citep{Cowie1977}, ``strangulation'' -- i.e. the cessation of cold gas replenishment due to the removal of hot circumgalactic gas \citep{Larson1980,Kawata2008}, viscous stripping of cold gas due to the flow of hot gas past the galaxy \citep{Nulsen1982}, and tidal effects due to interactions with other galaxies in the cluster or the cluster potential \citep{Moore1996,Moore1998,Smith2015}. 

The effect of these processes on the neutral atomic hydrogen  ($\HI$) of galaxies is well documented \citep[e.g.][]{Yun1994,Verdes2001,Kenney2004,Cortese2010,Yoon2017,Dudzar2019,Reynolds2021,Lopez2022}. Cluster galaxies, in particular, have been shown to have lower $\HI$ content
\citep{Giovanelli1985,Solanes2001,Cortese2011,Brown2017}, severely truncated $\HI$ disks \citep{Warmels1988,Bravo2000}, and disturbed $\HI$ morphologies \citep{Chung2009,Fumagalli2014}. However, there is no overall consensus regarding the impact of cluster environment on the molecular gas content of galaxies. This is despite a plethora of centimeter and (sub)millimeter observations of galaxies carried out after the advent of telescopes like the Atacama Large Millimeter/Submillimeter Array \citep[ALMA; ][]{Wootten2009}, Jansky Very Large Array \citep[VLA; ][]{Perley2011}, and the IRAM NOrthern Expanded 
Millimeter Array [NOEMA; an extension of the original Plateau de Bure interferometer (\citealt{Guilloteau1992})]. 

Many early as well as recent studies have claimed that galaxies in clusters contain similar \citep[e.g.][]{Casoli1991,Boselli1995,Rudnick2017,Darvish2018,Wu2018,Lee2021}, or even more molecular gas \citep{Noble2017,Hayashi2018,Tadaki2019}, than those not in clusters. On the other
 hand, there is also strong support for a deficit of molecular gas in such systems
 \citep[e.g.][]{Vollmer2008,Jablonka2013,Wang2018,Sperone2021,Alberts2022}. However, the interpretation of these results is not so straightforward. Galaxies detected in molecular gas observations must possess a decent amount of $\Hmol$ in the first place, and therefore, tend to be star-forming (hereafter SF), while the most affected systems are deficient in $\Hmol$, form a larger fraction of the cluster population, and evade detection. As such, statistically significant samples of SF systems in clusters spanning a wide range in cluster mass are required for conclusive results. Observational studies focussed on clusters, though, generally suffer from poor statistics and inhomogeneous sampling, which can introduce unintended biases and lead to conflicting results.
 
 It is true that some studies focussing on nearby clusters have shown direct evidence of ram-pressure stripping \citep{Vollmer2008,Vollmer2009,Zabel2019,Moretti2020}, but they disagree on its impact on the molecular gas content. This is partially because, before cluster galaxies begin to lose
 molecular gas, they may gain it due to, for instance, momentary compression of galactic gas by the ram pressure \citep{Fujita1999,Kronberger2008,Kapferer2009,Bekki2014,Henderson2016,Mok2016,Steinhauser2016,
 Lee2017,Martinez2018,Vulcani2018,Safarzadeh2019,Troncoso2020,Roberts2021,Lee2022}. On long timescales, 
 however, the molecular gas content is expected to reduce. Hence, different results regarding the $\Hmol$ 
 content of such galaxies can be obtained depending on the evolutionary stage that is observed. Moreover, 
 since $\Hmol$ is predominantly concentrated in the inner regions of a galaxy \citep[e.g. see][]{Sun2020}, where the gravitational 
 potential is deepest, it is rare to find cases where there is ongoing stripping of $\Hmol$, and big samples 
 are required in order to derive robust conclusions. 

 Furthermore, it is important to consider that galaxy populations in clusters differ between redshifts.
 \citet{Butcher1978}, and many others thereafter \citep{Balogh1999,Poggianti2006,Saintonge2008,Li2009,Brodwin2013,Raichoor2012,Hennig2017,Wagner2017,Jian2018,Pintos2019}, showed that clusters at high redshift have a higher fraction of blue SF galaxies compared to similar-mass clusters at $z=0$. The morphology--density relation in clusters also gets shallower at high $z$ \citep[e.g.][]{Postman2005}, and galaxies in cluster cores sometimes show higher SFRs than the field \citep[e.g.][]{Tran2010}. Taken together, these trends are attributed to the fact that clusters at high redshifts are newly formed and expected to go through an initial merger-driven phase, accompanied by the accretion of ample cool gas to support star formation. These factors, combined with poor statistics, complicate any investigation of the impact of environment on galaxies in high-$z$ clusters.

Cosmological hydrodynamical simulations can help circumvent the issues present in observational studies by providing sufficient statistics, complete samples of galaxies, and a diverse range of environments. More importantly, such simulations enable a proper investigation of temporal variation in galaxy properties, and its relation to external processes. For example, \citet{Stevens2021} used the IllustrisTNG simulations \citep{Pillepich2018,Nelson2018,Springel2018,Marinacci2018} to extensively explore the impact of environment on the $\Hmol$ content of $z=0$ galaxies. They found that the
$\Hmol$ content of satellites is, on average, lower than centrals by $\approx 0.6$~dex. They also showed that the mean $\Hmol$ content of satellites in haloes with masses above $10^{14}\,{\rm M_\odot}$ is lower by $\approx 1$~dex relative to satellites in haloes with masses less than $10^{12}\,{\rm M_\odot}$. This was attributed to mass loss post infall, which occurs for both $\HI$ and $\Hmol$, but at a faster rate for the former.

In this paper, we use the 100 Mpc ``REFERENCE'' box of the \eagle suite of hydrodynamical simulations \citep{Schaye2015,Crain2015} to investigate whether SF galaxies in clusters differ in their $\Hmol$ content in comparison to other SF galaxies, and what is the underlying physics governing this difference, or lack thereof. This work complements \citet{Stevens2021} and extends it out to $z=1$, roughly corresponding to half the age of the Universe. The paper is structured as follows. We describe the simulation and the methodology in Section~\ref{simandmethod}, where Section~\ref{sim} delineates details about the simulation data used in this work, and Section~\ref{galselect} explains the galaxy selection. In Section~\ref{propenv}, we compare the $\Hmol$ content of SF galaxies in clusters with those outside clusters, comment on the
factors that modulate the differences, and discuss the implications for galaxy quenching in clusters.
We then extend the analysis on SF cluster galaxies in Section~\ref{trackparts}, where we demonstrate the influence from various modes of molecular gas transfer on the $\Hmol$ content with respect to their infall, and also assess the contribution from feedback. We summarise our main findings and some important caveats of the study in Section~\ref{summary}. Additionally, for the interested reader, we describe 
our approach for modelling the $\Hmol$ in Appendix~\ref{postproc}, and the tests performed for identifying the valid prescriptions for \eagle in Appendix~\ref{prestest}.

\section{Simulation and Methods}\label{simandmethod}
\subsection{The \textsc{\large eagle} simulation}\label{sim}
Evolution and Assembly of GaLaxies and their Environments, or \eagleNS\footnote{\url{https://eagle.strw.leidenuniv.nl/}} \citep{Schaye2015,Crain2015}, is a suite of hydrodynamical simulations that were evolved with smoothed-particle hydrodynamics using a modified version of the \textsc{\large gadget-3} code (a successor to the \textsc{\large gadget-2} code described in \citealt{Springel2005}). Here, we briefly describe the salient aspects of the simulation pertinent to the analysis in this paper. For a more generalised description of \eagleNS, the interested readers should refer to \citet{Schaye2015}, \citet{Crain2015}, and the references therein.

In this work, we use the ``REFERENCE'' \eagle simulation corresponding to the box with a comoving side length $L=100\,{\rm Mpc}$ and $N=2\times 1504^3$ particles (including both dark matter 
and baryons). The simulation is based on a flat $\Lambda$CDM cosmology with cosmological parameters 
from the \citet{Planck2014} results.
Each dark matter and 
(primordial)\footnote{As the simulated universe evolves, the mass of a gas particle can differ 
from its primordial value owing to feedback processes that transfer mass into or 
out of the particle.} gas particle has mass $m_\text{dm} = 9.70 \times 10^6\,{\rm M}_\odot$ and $m_\text{g} = 1.81\times 10^6\,{\rm M}_\odot$, respectively. \eagle lacks the resolution required for modelling fine-scale processes, which are rather implemented using the subgrid physics described in \citet{Crain2015}. This, however, does not include the physics of the cold ISM. We compute the $\Hmol$ content for 
gas particles in post-processing, i.e. after the simulation was run. To this end, we first derive the neutral hydrogen content (Appendix~\ref{neuthyd}) using
the prescription of \citet{Rahmati2013}, and then subsequently partition it into atomic and molecular components (Appendix~\ref{molfrac}). For the latter, we test five different prescriptions that have been widely used in the literature, namely: \citet{BR06}, \citet{L08}, \citet{K13}, \citet{GK11}, and \citet{GD14}; hereafter BR06, L08, K13, GK11 and GD14, respectively. The description and implementation is provided in Appendix~\ref{molfrac}. We compare the outputs from these prescriptions against observational data and infer that only K13 and GD14 are sufficiently reliable for \eagle (see Appendix~\ref{prestest}). The results in this paper are based on GD14.\footnote{Though we only show the results for GD14, we have confirmed that similar trends exist for K13.}

Dark matter haloes in the simulation are first identified using a friends-of-friends (FoF) algorithm \citep{Davis1985}, and the star, BH and gas particles are considered to be part of the FoF halo containing the nearest dark matter particle, if the particle belongs to one. Each FoF halo along with its baryons is a `group',
and the self-bound substructures (or subhaloes) within each group are identified using \textsc{\large subfind} \citep{Springel2001,Dolag2009}. We consider the virial mass of a group as $M_{200}$, which is the total mass enclosed within the radius where the mean density is $200$ times the critical density of the universe. We define clusters as the groups with $M_{200}>10^{13.8}\,{\rm M}_\odot$ (see Section~\ref{propenv}). The most massive subhalo in each group is referred to as the `central' subhalo (which, if massive enough, may host the central galaxy), and the remaining subhaloes which orbit the central are `satellite' subhaloes and may host the satellite galaxies. For each subhalo -- and the corresponding galaxy -- the centre is considered to be the position of the particle with the minimum gravitational potential according to \textsc{\large subfind}. The centre of the central subhalo/galaxy is considered to be the centre of the group. The merger tree for each halo constitutes the complete set of branches of its surviving subhaloes \citep{Qu2017}, and we use them to construct histories for several properties of dark matter haloes and the galaxies therein.



\subsection{Selecting star-forming galaxies}\label{galselect}
We intend to adopt a selection strategy that results in populations that are (or close to) comparable to the heterogeneous observed samples across redshifts. As stated earlier, observed galaxies with molecular gas detections, especially at high redshifts ($z>0.5$), tend to be SF or star-bursting. This is unsurprising, since the systems that emit enough flux to allow for their detection by current telescopes are likely to be rich in molecular gas, and thus also have high SFRs. Considering this, we focus our analysis only on SF galaxies in \eagleNS.

\begin{figure*}
\begin{center}
  \includegraphics[width=2.1\columnwidth, trim={0.1cm 0.1cm 0.1cm 0.1cm}]{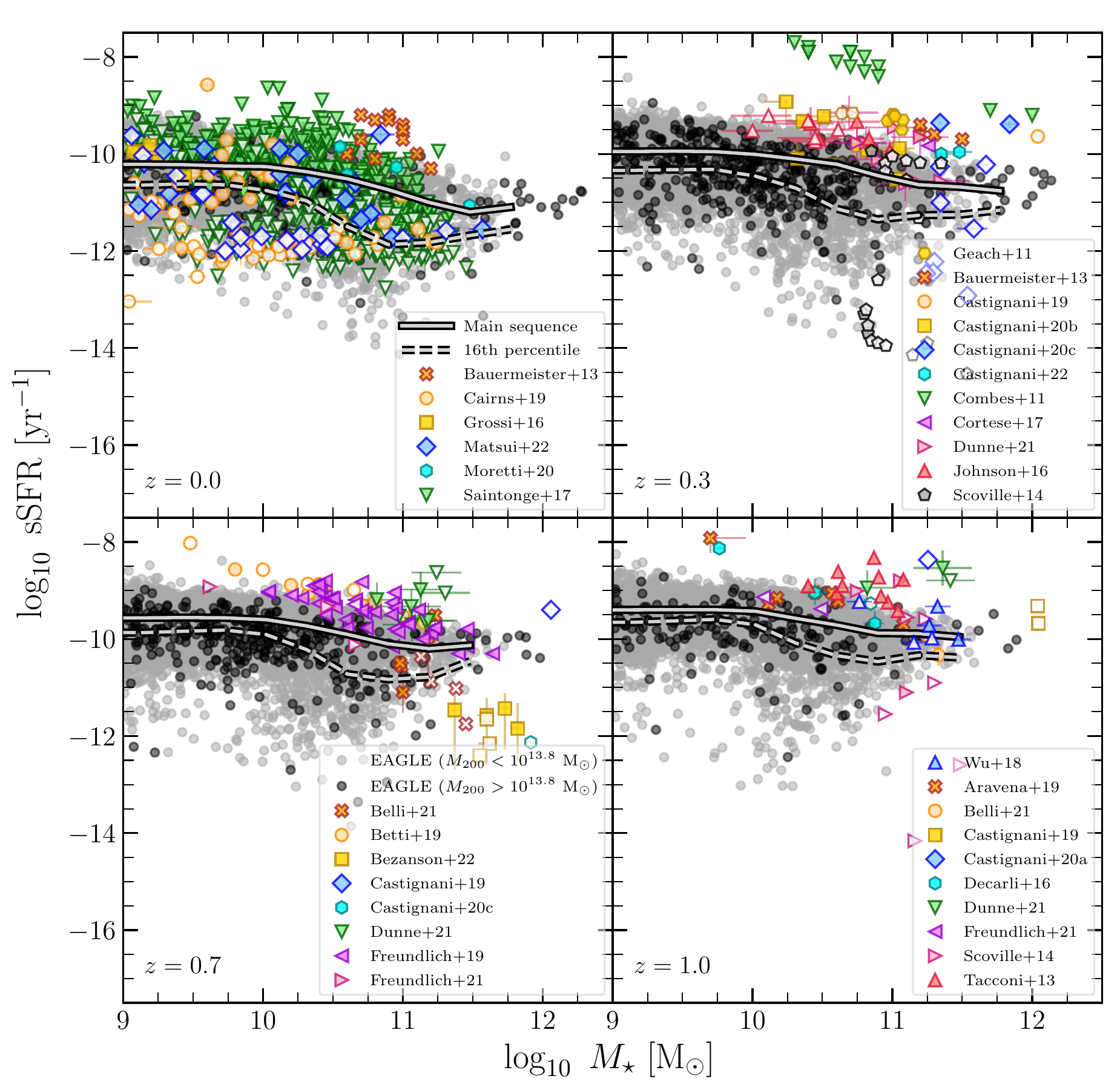}
   \caption{Specific star-formation rate plotted against stellar mass for \eagle galaxies, and for various observational datasets. For the latter, we only show galaxies that have molecular gas measurements. Each panel corresponds to a particular redshift between $z=[0,1]$ quoted in the bottom-left corner. The grey and black points are individual \eagle galaxies in haloes with $M_{200}<10^{13.8}\,{\rm M}_\odot$ and $M_{200}>10^{13.8}\,{\rm M}_\odot$, respectively. The filled symbols and open symbols correspond to galaxies with \textit{molecular gas} detections and upper limits. The solid grey curve shows the median sSFR or the ``main sequence'' in \eagle, 
   while the dashed grey curve shows the 16th percentiles for 12 bins of stellar mass spread across $M_\star/{\rm M}_\odot=[10^9,10^{12.5}]$; the curves
   are only shown for bins that have $>10$ galaxies. Most of the detections lie above the dashed curve at each of the four redshifts. For this study, we select all the \eagle galaxies above the 16th percentiles of sSFR.}
   \label{sampsel}
\end{center}
\end{figure*}

The star formation activity of a galaxy is usually characterized based on its position relative to the star formation main sequence, which is the average SFR--$M_\star$ or sSFR--$M_\star$ relation (sSFR is the specific star-formation rate or SFR/$M_\star$). We use the sSFR--$M_\star$ plane obtained for \eagle to select SF galaxies at four different redshifts: $z=0,0.3,0.7$ and $1$. Each panel in Fig.~\ref{sampsel} is the sSFR vs $M_\star$ scatter plot for \eagle and observed galaxies at the redshift
$\pm 0.1$ of the redshift written in the bottom-left corner are shown. The latter only includes the galaxies with molecular gas measurements: \citet{Geach2011,Bauermeister2013,Tacconi2013,Scoville2014,Decarli2016,Grossi2016,Johnson2016,Saintonge2017,Wu2018,Aravena2019,Betti2019,Cairns2019,Castignani2019,Freundlich2019,Castignani2020a,Castignani2020b,Castignani2020c,Moretti2020,Belli2021,Dunne2021,Freundlich2021,Bezanson2022,Castignani2022,Matsui2022}. We omit \eagle galaxies with $M_\star<10^9\,{\rm M}_\odot$, since the observed masses,
particularly at high redshifts, tend to lie above this threshold. This criterion also happens to avoid galaxies that are not reasonably resolved in \eagle \citep{Schaye2015}. The solid curve shows the main sequence (medians) at each redshift obtained for \eagle galaxies, and the dashed curve shows the 16th percentiles of the sSFR for 12 bins of stellar mass between $M_\star/{\rm M}_\odot=[10^9,10^{12.5}]$ (the galaxies with ${\rm sSFR}=0\,{\rm yr}^{-1}$ are excluded while calculating the percentiles). The curves are only shown for the bins with $>10$ galaxies. The observed galaxies with molecular gas detections and upper limits are shown as filled and empty symbols, respectively.

\begin{table*}
    \centering
    \begin{tabular}{c|c|c|c|c}
    \hline
    $z$ & SF / NSF & Cluster / Non-cluster & Cen / Sat (Cluster) & Cen / Sat (Non-cluster) \\
        &        & (SF)                & (SF)              & (SF) \\    
    (1) & (2) & (3) & (4) & (5)\\
    \hline
    0    & 8739 / 4561  & 263 / 8476 & 13 / 250 & 6140 / 2336 \\
    0.3  & 9936 / 3743  & 286 / 9650 & 14 / 272 & 6762 / 2888 \\
    0.7  & 10583 / 2725  & 190 / 10393 & 8 / 182 & 7212 / 3181 \\
    1.0  & 10297 / 2340  & 94 / 10203 & 4 / 90 & 7043 / 3160 \\
    \hline
    \end{tabular}
    \caption{The statistics of galaxies in \eagle for the four redshifts explored in this paper. The numbers are for galaxies with stellar masses $M_\star>10^{9}\,{\rm M}_\odot$. (1) lists the sample redshift; (2) lists the number of SF and non-SF galaxies (including ${\rm sSFR}=0\,{\rm yr}^{-1}$ systems); (3) lists the number of SF galaxies in clusters and in non-clusters; (4) lists the number of SF centrals and satellites in clusters; and (5) shows the number of SF centrals and satellites in non-clusters.}
    \label{galstats}
\end{table*}

Recall that each \eagle galaxy is part of a gravitationally bound substructure identified using \textsc{\large subfind}, and this is in turn located within a group which may also contain other galaxies. For example, in a system with a central galaxy surrounded by multiple satellite galaxies, each galaxy is surrounded by its own DM subhalo, and the whole system is inside a group with virial mass of $M_{200}$. We define `clusters' as those \eagle \textit{groups} that have mass $M_{200}>10^{13.8}\,{\rm M}_\odot$, and the rest as `non-clusters'. This threshold is somewhat arbitrary, as the transition mass is not a well-defined quantity. We adopt this value simply because it (approximately) delineates the lower bound of (proto-)cluster masses in molecular gas observations \citep[e.g.][]{Wagg2012,Tadaki2014,Casey2016,Noble2017,Castignani2018,Coogan2018,Strazzullo2018,Wang2018,Zabel2019,Alberts2022,Castignani2022}. The clusters thus selected have masses $M_{200}/{\rm M}_\odot \lesssim 10^{14.6},10^{14.4},10^{14.3},10^{14.1}$ at $z=0,0.3,0.7,1$, respectively. Fig.~\ref{sampsel} shows the \eagle galaxies residing in clusters as black points, and the others as grey points. 

The results show that, at each redshift, detections in the observational datasets predominantly lie on or above the dashed curve. Therefore, for each of these four redshifts, we select 
galaxies from \eagle that have sSFR above the 16th percentile \textit{for their stellar mass}.\footnote{This is done by interpolating the dashed-grey curve at the $M_\star$ of the galaxy, and selecting it if its sSFR is greater than the interpolated value.} The breakdown of galaxies between SF and non-SF (including ${\rm sSFR}=0\,{\rm yr}^{-1}$) for $M_\star>10^{9}\,{\rm M}_\odot$ is shown in the first column of Table~\ref{galstats}. Since ${\rm sSFR}=0\,{\rm yr}^{-1}$ galaxies do not constitute a substantial fraction, SF galaxies account for $\gtrsim 65$ per cent of the sample at each of the four
redshifts. The second column of Table~\ref{galstats} shows the number of SF galaxies that reside in clusters, and those that do not. Only $\lesssim 3$ per cent of all the SF galaxies in our sample are located in clusters. The last row shows that the number of SF cluster galaxies is $<100$ for $z=1$, and we find it to reduce further for higher redshifts. This is the reason for limiting our analysis to $z\leq 1$.

We realise that our adopted criteria are far from perfect if we are to match the observations
exactly. Even if we were to attempt that, it would not be accurate because the observed samples are heterogeneous and a single selection function cannot mimic the full dataset. In addition, the maximum sSFR of \eagle galaxies is almost an order of magnitude below some observations (Fig.~\ref{sampsel}), which is -- in part -- due to
the limited volume of \eagleNS, which does not sample enough rare, star-bursting galaxies. However, we would like to emphasise that an exact match with observations is \textit{not} our intention here. Instead, the aim is to exclude quiescent galaxies while also maintaining a sample that is fairly representative.\footnote{There are approaches alternative to ours that can be adopted for selecting SF galaxies. For example, a common approach is to select galaxies that lie above a fixed offset from the main sequence. We, however, do not adopt this approach because SF galaxies may not populate a region with a constant width around the main sequence for all stellar masses.}

Also, while there is no explicitly enforced limit on the gas content, the selection of SF galaxies inherently means that they must possess ample cold gas to sustain their star formation. Note that imposing a strict limit would not be appropriate if we want our selection to be guided by observations, as detections and upper limits overlap in the $M_\Hmol$--$M_\star$ space; this can be seen, for example, in Fig.~\ref{mh2xcold} for xCOLD GASS. Moreover, the strategies for computing upper limits differ between surveys.

We note that \textsc{\large subfind} occasionally identifies dense star-clusters within galaxies as subhaloes \citep{Mcalpine2016}. These structures have little stellar mass ($M_\star<10^8\,{\rm M}_\odot$), but anomalously high metallicity or BH mass. Our sample is devoid of such spurious subhaloes due to the adopted stellar mass cut.

\section{Molecular gas content of SF galaxies in clusters and non-clusters}\label{propenv}
\noindent In this section, we address the main aim of this study: 
the impact of cluster environment on the molecular gas content of SF galaxies, and its
dependence on redshift.
For this, at each redshift, we compare the SF \eagle galaxies that are located within clusters against those
that are not. Each of these two samples contains both central and satellite galaxies; the exact numbers are given in Table~\ref{galstats}. The number of clusters containing these galaxies is equal to the number of centrals, except for $z=0$ where the central in one of the clusters is not SF -- that is, there are actually 14 clusters with SF galaxies at this redshift. Satellites are generally considered to be more susceptible to environment than centrals, but we do not show results for the two classes separately here for the sake of consistency with observational studies, which cannot reliably distinguish between them in most cases. Notwithstanding, we do mention the satellite fractions if required for proper interpretation of the results. Overall, our selection causes the non-cluster sample to be predominantly composed of centrals ($\approx 70$ per cent), and the cluster sample to be dominated by satellites ($\approx 95$ per cent), for all redshifts. Note that each cluster galaxy below $M_\star\approx 10^{11.5}\,{\rm M}_\odot$ is a satellite, which is expected for haloes at this mass scale.

\begin{figure*}
\begin{center}
\includegraphics[width=2\columnwidth,trim={0.1cm 0.1cm 0.1cm 0.1cm}]{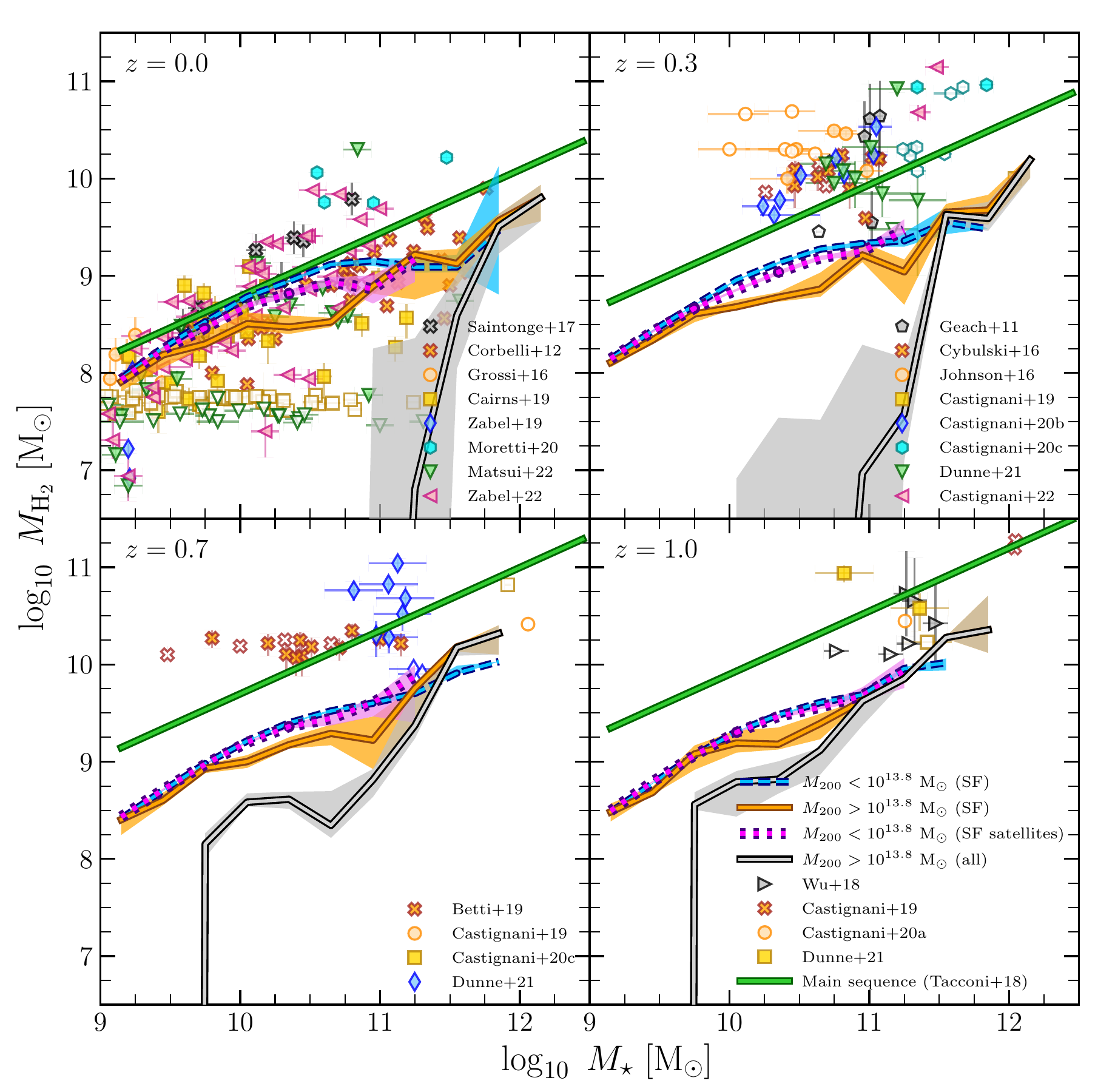}
   \caption{$\Hmol$ mass plotted against stellar mass for SF galaxies in clusters and
   non-clusters. The solid orange and dashed blue curves with outlines are the medians based
   on the GD14 prescription for the cluster sample and the non-cluster sample, respectively. The dotted magenta curve is the median for satellite galaxies
   in the non-cluster sample. The solid grey curve is the median for all the cluster galaxies. The curves
   are constructed using 0.3~dex wide stellar-mass bins spanning $M_\star/{\rm M}_\odot=[10^9,10^{12.3}]$. The shaded regions show the $1\sigma$ bootstrapping uncertainties of the medians. The filled and open squares show the detections and upper limits for the observed galaxies in (proto-)clusters, respectively. The solid green line is the relationship for
   galaxies on the observed star-formation main sequence from \citet{Tacconi2018}.
   Each panel corresponds to the redshift labelled in the top-left corner. The $\Hmol$ content of galaxies in clusters is generally lower than galaxies with the same stellar mass that do not reside in clusters. The low-mass SF galaxies with $M_\star\lesssim 10^{9.3}\,{\rm M}_\odot$ do not exhibit this offset. This also seems true for $M_\star\gtrsim 10^{11.2}\,{\rm M}_\odot$, but the statistics are poor in this regime ($<10$ galaxies per bin). At a fixed stellar mass, the offset between the two samples, on average, increases as the redshift decreases.}
   \label{mh2vsmstar_allz}
\end{center}
\end{figure*}

In Fig.~\ref{mh2vsmstar_allz}, we show the $M_{\Hmol}$--$M_\star$ relations for the two samples at each of the four redshifts, along with observations. The median relations for \eagle (based on GD14) are shown as the outlined solid orange curve for SF galaxies in clusters ($M_{200}>10^{13.8}\,{\rm M}_\odot$) and dashed blue curve for non-clusters ($M_{200}<10^{13.8}\,{\rm M}_\odot$); using 0.3~dex wide bins that lie within $M_\star/{\rm M}_\odot=[10^9,10^{12.3}]$. The solid grey curve shows the medians computed if we consider all the cluster galaxies. The shaded regions with the same colour coding show $1\sigma$ bootstrapping errors on the medians, i.e. they span the 16th and 84th percentiles of the medians of the bootstrap samples. The dotted magenta curve shows the median based on satellites in the non-cluster sample.\footnote{These are calculated using the {\tt bootstrap\_percentiles} module at \url{https://github.com/arhstevens/Dirty-AstroPy/blob/master/galprops/galcalc.py}.} The filled and open squares show the detections and upper limits, respectively, for galaxies in (proto-)clusters taken from \citet{Geach2011,Corbelli2012,Cybulski2016,Grossi2016,Johnson2016,Saintonge2017,Wu2018,Betti2019,Cairns2019,Castignani2019,Zabel2019,Castignani2020a,Castignani2020b,Castignani2020c,Moretti2020,Dunne2021,Castignani2022,Matsui2022,Zabel2022}. For systems that are redundant between the datasets, the most recent values are shown. The scaling relation for the observed star-formation main sequence by \citet{Tacconi2018} is displayed as the
solid green line, and is based on a representative sample of SF galaxies between $z=0-4.4$. 
This relation is applicable to molecular gas masses based on both CO and dust emission.

The $z=0$ results for \eagle indicate that, except for the lowest and highest stellar masses, SF galaxies in clusters typically have a deficit of $\Hmol$ compared to their non-cluster counterparts. The difference -- at $\approx 0.5$~dex -- is most pronounced for $M_\star \approx 10^{10.5}\,{\rm M}_\odot$. The trend also exists at $z>0$, albeit with offsets that get smaller with increasing redshift at a given stellar mass. We find that, at each redshift, there are certain stellar mass bins where differences between the medians are significant even after accounting for their bootstrapping errors. Hence, these results establish that (simulated) SF galaxies in clusters have lower $\Hmol$ content than those not in clusters, provided they lie in a certain mass range.

The situation, though, is not so clear for observed galaxies, and the data shown in Fig.~\ref{mh2vsmstar_allz} help
demonstrate why that is the case. In observations, the impact of the cluster environment is usually inferred by comparing the molecular gas content of cluster galaxies to that of galaxies on the main sequence at the cluster's redshift. If we assume that the main-sequence relation from \citet{Tacconi2018} represents non-cluster galaxies (which is a reasonable
assumption because they dominate the statistics), we can infer an enhancement or deficit of molecular gas in clusters depending on the dataset to which we choose to compare. For example, at $z=0$, 
all of the \citet{Moretti2020} detections lie above the observed main sequence, whereas most of the data from \citet{Cairns2019} lie below it. To add to this, an accurate determination of the difference in the $\Hmol$ content using observational data is limited by the presence of non-detections, which do not contaminate the simulated data.

Fig.~\ref{mh2vsmstar_allz} also shows that SF cluster galaxies are not representative of the whole cluster population. For any stellar mass bin with reliable statistics ($>10$ galaxies), an average cluster galaxy contains less $\Hmol$ than a typical SF one, and this difference
increases at lower masses, indicating that SF systems sample a smaller proportion of the population. This is somewhat expected, given that susceptibility to environment increases as a galaxy's mass decreases, which then reduces the likelihood of preserving the cold gas reservoir \citep[e.g.][]{Marasco2016}. We predict that, for the cluster masses explored here, observations carried out using a radio/IR telescope with sufficient sensitivity to detect gas-poor systems would produce a $M_\Hmol$--$M_\star$ relation that is (at least qualitatively) similar to the solid grey curve.

Note that the difference in $\Hmol$ content of the cluster and non-cluster SF samples could be a representation of the difference in their satellite fraction, which is evidently greater for the former (Table~\ref{galstats}). We investigate this by plotting the median $\Hmol$ content of satellites in the non-cluster sample in Fig.~\ref{mh2vsmstar_allz} (dotted curve). The offsets still persist relative to the full cluster sample, and at similar magnitudes as for the full non-cluster sample. Therefore, the offsets between the cluster and non-cluster sample are also indicative of differences between the satellites in these two kinds of groups, and not simply driven by the difference in satellite fraction.

\begin{figure*}
\begin{center}
  \includegraphics[width=1.7\columnwidth,trim={0.5cm 0.1cm 0.1cm 0.1cm}]{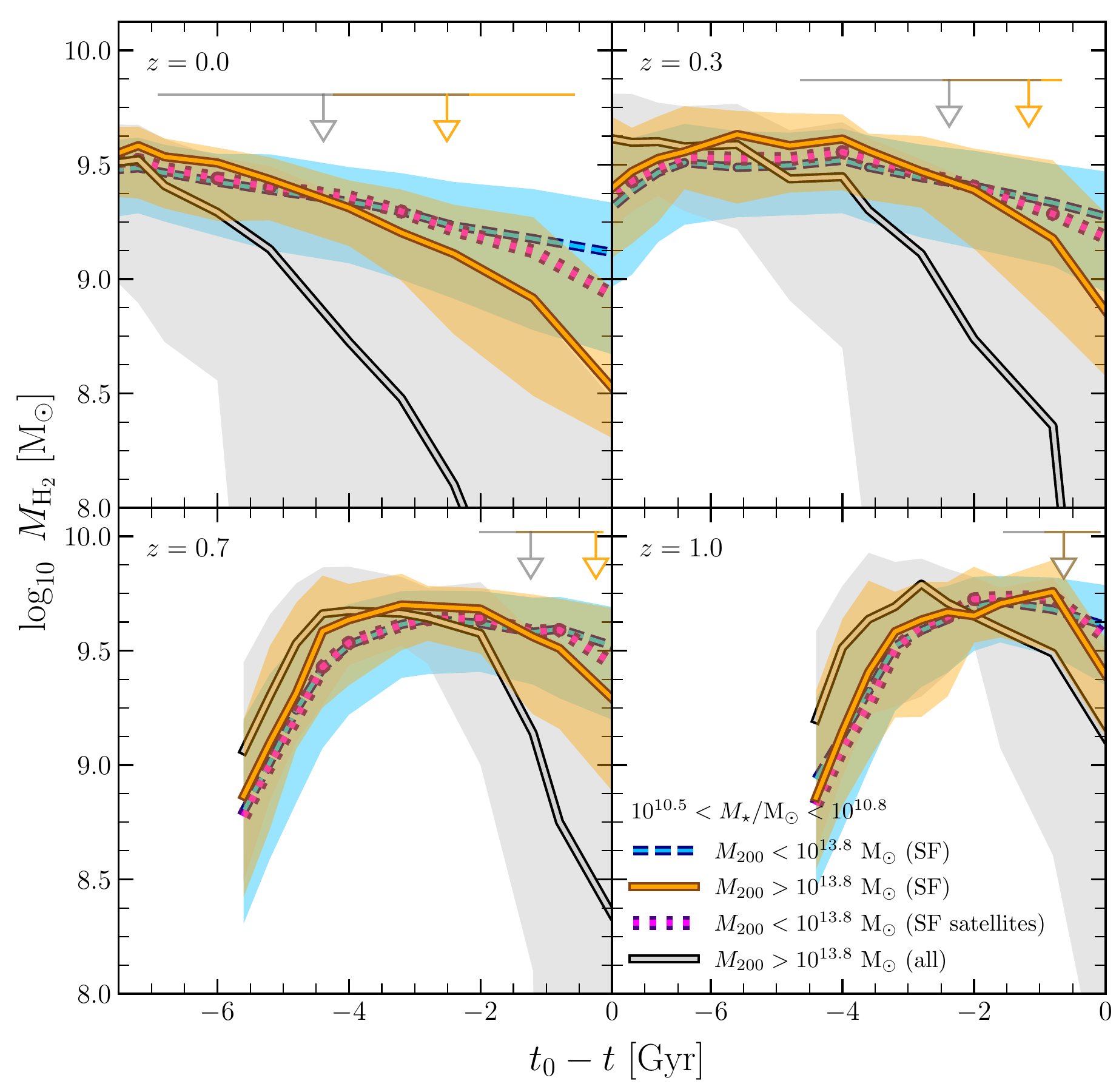}
   \caption{The $M_\Hmol$ history for the main progenitors of galaxies with $10^{10.5}<M_\star/{\rm M}_\odot<10^{10.8}$. Each panel corresponds to the redshift written in the top-left corner. $t$ is the look-back time corresponding to the progenitor's redshift, and $t_{\rm 0}$ is the look-back time at the sample redshift; cosmic time moves forward from left to right on the $x$-axis. The curves in each panel show the median histories and follow the same plotting style and colour coding as in Fig.~\ref{mh2vsmstar_allz}. The corresponding shaded regions enclose the 20th to 80th percentiles. The downward arrows indicate the median value of the time of first infall into a cluster, $t_{\rm c}$, for the cluster sample -- i.e. the look-back time when they are bound to a group with $M_{200}=10^{13.8}\,{\rm M}_\odot$. The corresponding error-bars span the 20th to 80th percentiles. This shows that cluster galaxies had similar $\Hmol$ content in the past, meaning that the galaxies outside clusters did not start off with particularly more $\Hmol$. At a moment approximately coinciding with $t=t_{\rm c}$, there was a sharp decline of the SF cluster sample relative to their non-cluster counterparts, which explains the lower $\Hmol$ content of cluster galaxies at $t=t_{\rm 0}$. The full cluster sample shows a steeper decline, but unlike the SF subsample, it appears to have commenced prior to their infall into a cluster.}
   \label{mh2vst}
\end{center}
\end{figure*}

The main-sequence relation itself is broadly consistent with \eagle at $z=0$ except at $M_\star>10^{11}\,{\rm M}_\odot$ (dashed blue and
violet curves), but its normalisation gets progressively higher than that of \eagle as redshift increases. 
Although the exact reason is uncertain, there are a number of factors that can potentially contribute to this. One possibility is that the detections are biased toward galaxies that are more gas-rich as redshift increases (Malmquist bias). Another is that the uncertainties regarding the CO-to-$\Hmol$ conversion factor can, in principle, lead to a systematic effect on the inferred masses. It may also reflect
the limitations of the post-processing schemes adopted to compute $\Hmol$ masses for \eagle galaxies at higher redshifts. Nevertheless, our focus is on understanding the offset in molecular gas content in different environments, which is more likely to be a function of the physics \emph{in the simulation} rather than any post-processing scheme. Thus, the results in Fig.~\ref{mh2vsmstar_allz} are useful.

We note that \eagle clusters contain more hot gas compared to observed clusters, due to AGN feedback not being efficient enough to eject the gas out of their host haloes \citep{Barnes2017}. Although this does lead to more massive centrals \citep[see][]{Bahe2017}, our cluster sample is dominated by satellites instead. \citet{Barnes2017} showed that the feedback in centrals is only strong enough to transport the gas out to $0.5\lesssim r/R_{500} \lesssim 1$, which means elevated density of intracluster gas, and therefore, stronger ram pressure acting on satellites at such radii. However, since the stellar mass function of cluster satellites in \eagle is consistent with observations \citep{Bahe2017}, this does not seem to have a significant impact on the global properties of
such systems. The exact physics involved remains elusive, and a proper investigation would require a detailed analysis, which is beyond the scope of this paper.

\subsection{Potential factors modulating the offset in $M_\Hmol$}\label{modfact}
Though the offset in $M_\Hmol$ is visible at all redshifts, its magnitude at a given stellar mass increases towards lower redshifts. The trend is particularly striking for the sample with all the cluster galaxies (solid grey curve, Fig.~\ref{mh2vsmstar_allz}). One potential reason for this could be that galaxies residing in clusters at lower redshifts have been bound to a cluster, and thus influenced by the cluster environment, for longer durations. We now investigate the validity of this scenario.

Each galaxy in \eagle has formed through mergers of smaller subhaloes, 
referred to as the galaxy's `progenitors'. The progenitors at a given snapshot along the most massive branch of its merger tree is called the `main progenitor' \citep[see][]{Qu2017}. We study the acquisition of our galaxies into their groups and its impact on the $\Hmol$ content by tracking their main progenitors, noting their masses and that of their host \textit{group}. For each cluster galaxy, we interpolate the group mass history to determine the look-back time ($t_{\rm c}$) when $M_{200}=10^{13.8}\,{\rm M}_\odot$, i.e. the \textit{first} instant it was bound to a cluster. 

Fig.~\ref{mh2vst} shows the $M_\Hmol$ histories for the cluster and non-cluster samples, where each panel corresponds to the redshift written in the bottom-right corner. $t$ is the look-back time corresponding to the progenitor's redshift, and $t_{\rm 0}$ is the look-back time at the sample redshift; cosmic time moves forward from left to right on the $x$-axis. The solid and dashed curves in each panel show the median $M_\Hmol$ history for the cluster and non-cluster samples, respectively. The corresponding shaded regions enclose the 20th to 80th percentiles. The dotted curve is the median for satellite galaxies in the non-cluster sample. The downward-pointing arrow is the median $t_0-t_{\rm c}$, corresponding to first infall into a cluster, and the horizontal
error bar shows its 20th--80th percentile range. Given that the $M_\Hmol$ offset
varies with $M_\star$, it is important to control for stellar mass for proper correspondence with the results in Fig.~\ref{mh2vsmstar_allz}. Here, we show the results for the galaxies with $10^{10.5}<M_\star/{\rm M}_\odot<10^{10.8}$ just as an example, but have verified that similar results apply to other stellar mass bins that show noticeable $\Hmol$ deficits.

These results show that, at some point in the past, both the samples possessed similar quantities of $\Hmol$, which provides an important insight: the galaxies outside clusters did not start off with more $\Hmol$ content to begin with. Later on, the cluster galaxies incurred a steep decline in their $M_\Hmol$ relative to the non-cluster sample, and this evolution eventually led to the offset seen at $t=t_0$. A similar history is seen for the full cluster sample, but note the steeper decline. The non-cluster sample with only satellites also shows
an average decline in $\Hmol$ mass relative to the full non-cluster sample, but -- as expected -- not as steep as the cluster samples. 

\begin{figure}
\begin{center}
  \includegraphics[width=1.0\columnwidth,trim={0.4cm 0cm 0cm 0.1cm}]{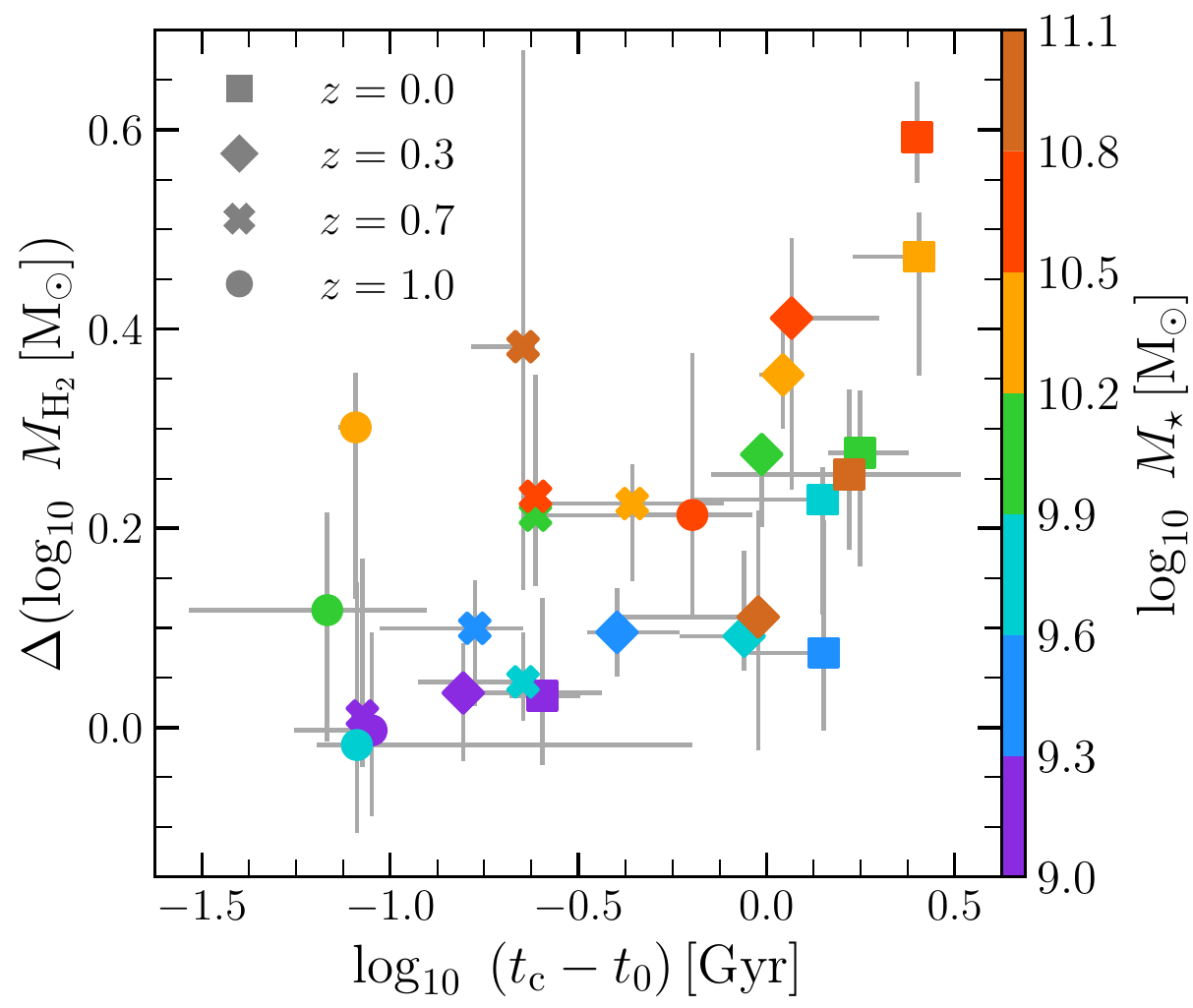}
   \caption{The offset between median logarithmic $\Hmol$ mass of SF galaxies in clusters and those not in clusters (i.e. the difference between the dashed blue and solid orange curves in Fig~\ref{mh2vsmstar_allz}), plotted against the median time elapsed since first infall of the cluster galaxies into clusters. $t_{\rm c}$ is the look-back time corresponding to the instant when a galaxy was first accreted by a cluster (group
   mass $M_{200}=10^{13.8}\,{\rm M}_\odot$), and $t_0$ is the look-back time corresponding to the sample redshift. Each data point pertains to a certain stellar mass bin and redshift, where the former can be gauged using the colour bar with labels as the bin edges used to construct the median curves in Fig.~\ref{mh2vsmstar_allz}, and the latter is conveyed through a specific symbol displayed in the top-left corner. For the sake of robustness, the results are shown only for stellar mass bins containing $>10$ galaxies in both the cluster and non-cluster samples, which excludes all the bins beyond $M_\star=10^{11.1}\,{\rm M}_\odot$. Overall, the $\Hmol$ offset is positively correlated to the time elapsed since infall. For a fixed stellar mass (colour), the offset generally reduces at higher redshifts (from square to circle), and the cluster galaxies have more recent infall times. For
   a given redshift (symbol), the $\Hmol$ offset -- on average -- decreases for progressively smaller stellar masses.}
   \label{dmh2vstsinf}
\end{center}
\end{figure}

Interestingly, the instant when the SF cluster sample deviated seems to have roughly coincided with the time when a cluster first accreted these galaxies. However, the wide error-bars on the arrows mean that the curves are averaging over a wide range of infall times, and are therefore not ideal for inferring the typical behaviour of a galaxy with respect to its infall into a cluster. Hence, we estimate the global significance of this trend by examining the quantitative relationship between the offset in $M_\Hmol$ and $t_{\rm c}-t_0$, or the time elapsed since infall. The
scatter plot is shown in Fig.~\ref{dmh2vstsinf}, where each data point shows the offset between the median logarithmic $\Hmol$ masses (underpinning the dashed blue and solid orange curves in Fig.~\ref{mh2vsmstar_allz}), and the median $t_{\rm c}-t_0$, for a particular stellar mass bin and redshift, depicted using a specific colour and symbol, respectively. The labels on the colour bar are the bin edges used to compute the medians, and are the same as
those used for Fig.~\ref{mh2vsmstar_allz}. We only
show the results for the bins that contain $>10$ galaxies, which effectively excludes all the bins beyond $M_\star=10^{11.1}\,{\rm M}_\odot$. The error-bars are the
bootstrapping uncertainties.\footnote{Since the lower and upper bootstrapping errors on the median $\Hmol$ are generally unequal, we derive the errors for the $\Hmol$ offset using the \citet{Barlow2003} method, implemented using the {\tt add\_asym} module at \url{https://github.com/anisotropela/add_asym/blob/master/add_asym.py} \citep[described and tested by][]{Laursen2019}.} 

The figure shows that the $\Hmol$ offset is indeed positively correlated with the time elapsed since infall for cluster galaxies; the Spearman rank coefficient and $p$-value 
for the whole sample is $\approx 0.57$ and $\approx 0.002$, respectively. The scatter at a fixed $t_{\rm c}-t_0$ 
is likely due to some additional factors. For example, a higher group mass and proximity to group's centre generally leads to a lower gas accretion rate \citep[e.g.][]{van2017}. For a fixed mass, group haloes at different redshifts have varied crossing times, which means that galaxies accreted at those redshifts have traversed a different number of orbits for the same time elapsed since infall. These haloes also have different intrahalo gas densities at their virial radius, which translates to different strengths of ram pressure for the same galaxy velocity. We assess the importance of potential modulating factors through a principal component analysis (PCA), which is a statistical technique that derives orthogonal vectors (called principal components) along which the variance is maximised in a multidimensional parameter space. This is utilised in galaxy evolution studies to find useful trends that may otherwise remain undiscovered due to noise in the data \citep[e.g.][]{Costagliola2011,Bothwell2016,Lagos2016,Cochrane2018}. For our purposes, we conduct the analysis\footnote{This was carried out using the \textsc{\large python} module at \url{https://scikit-learn.org/stable/modules/generated/sklearn.decomposition.PCA.html}} on the dataset containing the current group mass, gas density at $R_{200}$ at infall (taken as that of dark matter; $\rho_{\rm 200,inf}$), crossing timescale of group at infall ($t_{\rm cross,inf}$), ratio of the galaxy's mass to the group's mass at infall $(M_{\rm gal}/M_{200})_{\rm inf}$, $t_{\rm c}-t_0$, and the $\Hmol$ offset. Fig.~\ref{dmh2vstsinf} can be envisaged as a projection of this hyperspace on the plane defined by the $\Hmol$ offset and $t_{\rm c}-t_0$. 

\begin{table*}
    \centering
    \begin{tabular}{c|c|c|c|c|c|c|c}
    \hline
    {PC} & {Explained variance} & $\log_{10} M_{200}$ & $\log_{10} \rho_{\rm 200,inf}$ & $\log_{10} t_{\rm cross,inf}$ & $\log_{10} (M_{\rm gal}/M_{200})_{\rm inf}$ & $\log_{10} (t_{\rm c}-t_0)$ & $\Hmol$ {offset}\\
    \hline
    {1}  & {0.61} & {0.45}  & {-0.47} & {0.47} & {0.21} & {0.48} & {0.29} \\
    {2}  & {0.27} & {0.24} & {-0.26} & {0.32} & {-0.65} & {-0.17} & {-0.56} \\
    {3} & {0.06} & {-0.43} & {-0.27} & {0.17} & {0.55} & {0.11} & {-0.63} \\
    \hline
    \end{tabular}
    \caption{Results of the principal component analysis. Each row represents a principal component, which is a vector in the parameter space orthogonal to all the other principal components. The first column shows its rank, and the second column shows the fraction of the total variance that is explained by the component. Any other column shows the coefficient for a particular parameter
    for that component. This includes the current group mass ($M_{200}$), the intrahalo gas density at the virial radius at infall ($\rho_{\rm 200,inf}$), the crossing time
    of the group halo at infall ($t_{\rm cross, inf}$), the ratio of galaxy mass to group mass at infall $(M_{\rm gal}/M_{200})$, time since infall ($t_{\rm c}-t_0$), and the $\Hmol$ offset (on
    the y-axis of Fig.~\ref{dmh2vstsinf}).}
    \label{pca}
\end{table*}
All the parameters are standardised before performing the PCA to ensure that the significance of a quantity is agnostic to 
its dynamic range. The results from this analysis are presented in Table~\ref{pca}, where the first column states
the principal component's rank, the second column shows the percentage of the total variance of the data explained by the component, and the rest of the columns show the coefficients for the various
parameters in the dataset. The coefficient's absolute value indicates the level of importance of
the corresponding parameter for that principal component. Most of the data can be explained by the first two components, accounting for $\approx 88$ per cent of the variance. We focus on these two
for the following. 

The first component represents $\approx 61$ per cent of the variance, and the coefficients show that most of it is contributed by $M_{200}$, $\rho_{\rm 200,inf}$, $t_{\rm cross,inf}$ and $t_{\rm c}-t_0$. The similar coefficient magnitudes suggest significance of parameters other than $t_{\rm c}-t_0$, and the dominance of four parameters implies that the component is useful for understanding trends mainly amongst them. The $t_{\rm c}-t_0$ is positively correlated to $M_{200}$ and $t_{\rm cross,inf}$ but negatively correlated to $\rho_{\rm 200,inf}$. The latter hints towards the importance of ram pressure, as a galaxy loses cold gas quicker and sustains its star formation for shorter durations under stronger pressures. $\rho_{\rm 200,inf}$ and $t_{\rm cross,inf}$ have
the same strengths, partially because both quantities are just functions of the
critical density of the Universe.

The second component explains $\approx 27$ per cent of the variance and is primarily composed of $(M_{\rm gal}/M_{200})_{\rm inf}$ and the $\Hmol$ offset, with coefficients almost twice those for other parameters. The coefficients suggest that these quantities are positively correlated for some systems, which is contrary to the expectation if gas stripping dominates the offset, as its efficacy is anti-correlated to this ratio \citep[e.g.][]{Tormen2004,Bahe2015,Yun2019}. This can nevertheless 
be caused by feedback events through the proverbial ``mass quenching'', which is more effective at higher masses due to hotter
circumgalactic gas \citep[e.g.][]{Gabor2015}. We will investigate this further in later sections. Overall, these results showcase the parameters that are contributing to the scatter in Fig.~\ref{dmh2vstsinf}.

The correlation between $\Hmol$ offset and the time since infall is also apparent in Fig.~\ref{dmh2vstsinf} for points corresponding to a given redshift (same symbol) or stellar
mass (same colour). For the latter, note that $t_{\rm c}-t_0$ always decreases at higher $z$ \textit{regardless} of how the $\Hmol$ offset behaves for that stellar mass, because of less cosmic time elapsed since the cluster's formation due to hierarchical assembly of haloes,\footnote{For instance, if we take the formation time as the look-back time when the mass of the group that currently hosts the cluster galaxy reached $10^{13.8}\,{\rm M}_\odot$, the median
time elapsed since formation reduces from $\approx 7.89$~Gyr for $z=0$ to $\approx 0.64$~Gyr for $z=1$.}
and shorter quenching timescales for cluster galaxies at higher redshifts \citep[e.g.][]{Baxter2022}. In addition, we find that these galaxies were always centrals before infall ($t=t_{\rm c}$), meaning they were not pre-processed in groups. Overall, these results support the notion that the trends in Fig.~\ref{mh2vsmstar_allz} can be attributed to the impact of the cluster environment on galaxies. 

It is worth noting that the $\Hmol$ offset is non-existent at $M_\star\lesssim10^{9.3}\,{\rm M}_\odot$, which may seem peculiar at first glance, as one might expect such low-mass galaxies to be most susceptible to environmental processes due to their shallow gravitational potential wells. Even the $\Hmol$ histories of these galaxies are very similar to their counterparts outside clusters, as if their environment had no influence on them. We know that such low-mass satellites in
$z=0$ \eagle clusters generally get quenched by $\lesssim 1$~Gyr after infall \citep{Wright2019}. In fact, low-mass SF galaxies are rare in clusters, constituting $\approx 10$ per cent of all the low-mass cluster satellites at $z=0$ \citep[in agreement with][]{Shao2018}, and $\approx 35$ per cent at $z=1$. Moreover, the cluster galaxies at this $M_\star$ tend to be devoid of $\Hmol$, regardless of the redshift (see Fig.~\ref{mh2vsmstar_allz}). This implies that such galaxies can be SF while \textit{also} being in a cluster only if they were accreted recently, as they would not be able to preserve enough cold gas otherwise. As it turns out, all the SF cluster galaxies with $M_\star<10^{9.3}\,{\rm M}_\odot$ are satellites, and most of these have only been bound to a cluster for $\lesssim 300$~Myr.

Similarly, Fig.~\ref{mh2vsmstar_allz} shows that the $\Hmol$ content of SF galaxies with $M_\star\gtrsim 10^{11.2}\,{\rm M}_\odot$ is oblivious to the cluster environment. However, this cannot be simply attributed to a short time span post infall in all cases. For example, the $z=0$ cluster galaxies at $M_\star \approx 10^{11.2}\,{\rm M}_\odot$ are satellites with a median $t_{\rm c}-t_0$ of $\approx 2$~Gyr, which is 
similar to the quenching timescale for such systems \citep{Wright2019}. Though, considering that the efficacy of gas stripping decreases with the satellite--to--central mass ratio \citep[e.g.][]{Tormen2004,Bahe2015,Yun2019}, it is plausible for 
a few massive galaxies in clusters to protect their gas reservoirs against such mechanisms for durations longer than usual. At $M_\star\gtrsim 10^{11.4}\,{\rm M}_\odot$, the SF cluster galaxies are mostly centrals at all redshifts, and are therefore relatively immune to environmental stripping. We, however, would also like to qualify that the data suffers from poor statistics in this range.

\subsection{Implications for galaxy quenching in clusters}
While it is well accepted that infall into a cluster exacerbates quenching, there has been an ongoing debate on the details of how SFR declines post infall. Deriving an accurate model for this is important for a comprehensive understanding of galaxy evolution, because the rate of decline of a galaxy's SFR provides crucial insights for the dominant quenching mechanism in clusters; for instance, ram-pressure stripping causes a more abrupt decline than reducing gas accretion. 

In a seminal study, \citet{Wetzel2013} showed that observed properties of cluster galaxies (like colours and SFRs) cannot be explained by a model that only assumes rapid or slow quenching. In an attempt to resolve this,  they proposed the `delayed-then-rapid' quenching scenario, wherein a satellite's SFR remains unaffected by the environment for a few Gyr after infall, and then incurs a rapid drop. This idea has amassed significant support over the last decade \citep[e.g.][]{Mok2014,McGee2014,Muzzin2014,Fossati2017,Foltz2018,Moutard2018,Rhee2020,Oman2021,Sampaio2022}. \citet{Haines2015} purported a similar `slow-then-rapid' model -- involving an initial phase of gradually reducing SFR -- as a more accurate model, because of its utility in explaining the observations of galaxies transitioning between SF and passive. This model has recently been gaining advocacy \citep{Maier2019,Roberts2019,Kipper2021}, and is generally understood physically as the `slow' phase caused by strangulation, and the `rapid' phase corresponding to the galaxy reaching inner regions of its group halo (within $0.5R_{200}$) and experiencing strong cold gas stripping due to ram pressure and tidal forces. Futhermore, it has been suggested that the time spent in the `rapid' phase reduces with stellar mass \citep[e.g.][]{Rhee2020}, which is usually interpreted as feedback events being more important for massive galaxies. Therefore, it is warranted to investigate the evolution of our cluster samples in the broader context of galaxy quenching.

\begin{figure*}
\begin{center}
  \includegraphics[width=1.7\columnwidth,trim={0.5cm 0.1cm 0.1cm 0.1cm}]{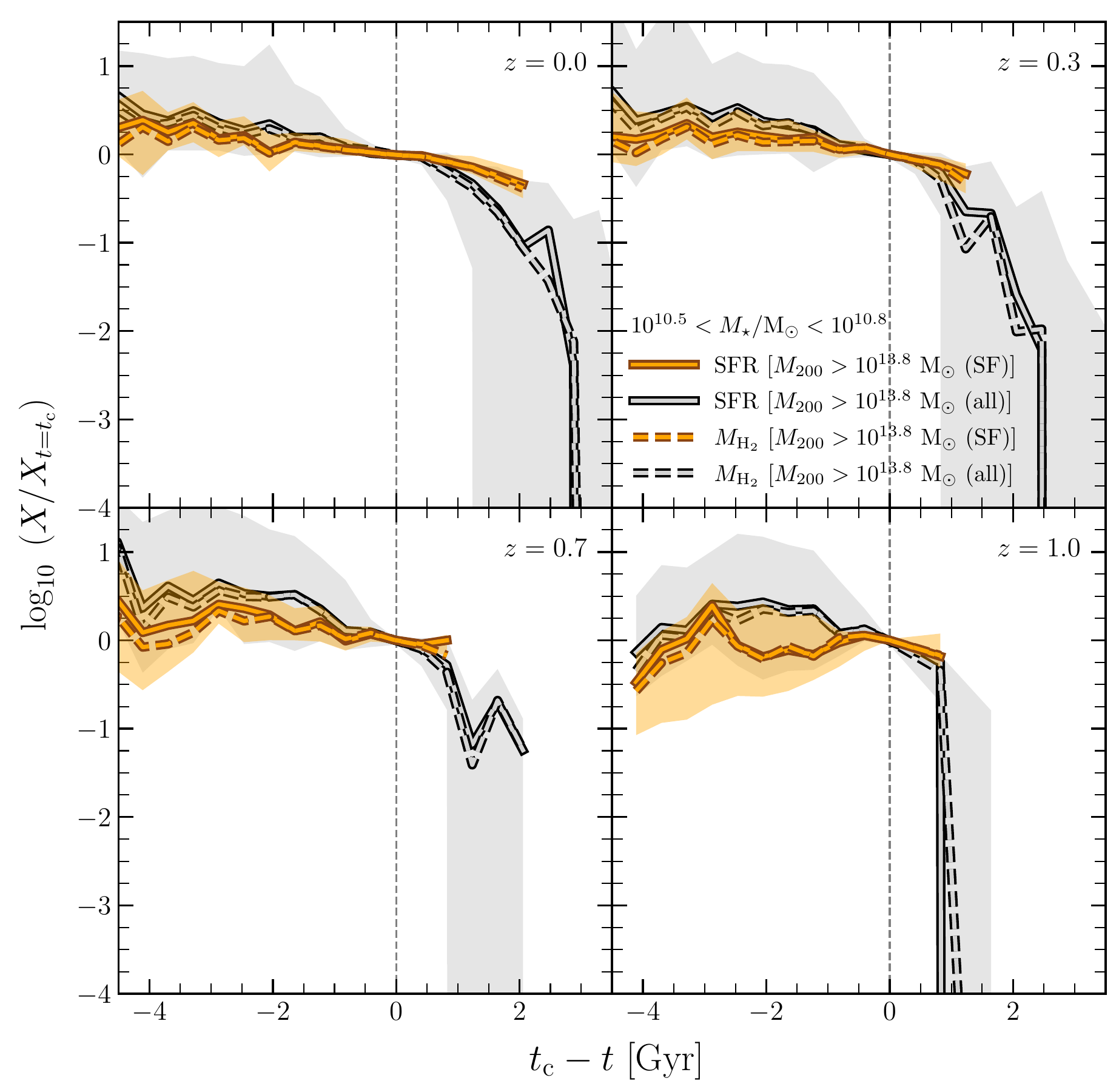}
   \caption{The evolution of SFR (solid curves) and $M_\Hmol$ (dashed curves) with respect to first infall into a cluster (corresponding to $t=t_{\rm c}$) for galaxies with $10^{10.5}<M_\star/{\rm M}_\odot<10^{10.8}$. Both quantities are normalised to their values at $t=t_{\rm c}$. The colour coding follows Fig.~\ref{mh2vst} and the shaded regions show the 20th to 80th percentiles for the SFR histories. The vertical dashed line shows the infall time. For the average cluster galaxy, the decline in SFR started about 2 Gyr before infall, and became stronger post infall. The top panels indicate that a typical evolution after infall involves an initial period of slow decline followed by a rapid drop in SFR, consistent with the so-called `slow-then-rapid' quenching paradigm. SF cluster galaxies experienced quenching only after infall and remained in the `slow' phase, with a decline rate that was weaker than that of an average cluster galaxy. The dashed curves show that 
   $M_\Hmol$ evolves in tandem with SFR, indicating close correspondence between molecular gas loss and quenching.}
   \label{xvstsi}
\end{center}
\end{figure*}

In Fig.~\ref{xvstsi}, we show the SFR and $M_\Hmol$ normalised by their values at infall for all the cluster galaxies (grey) and the SF subsample (orange) with $10^{10.5}<M_\star/{\rm M}_\odot<10^{10.8}$. The solid grey curves show that the SFR for a typical cluster galaxy started declining $\gtrsim 2$~Gyr prior to infall, and the decline got stronger from infall onwards. This indicates that the cluster environment expedites quenching of galaxies at all redshifts, confirming the findings from previous studies based on both observations \citep[e.g.][]{Chung2011,Haines2015,Pintos2019,Stevens2021}, and simulations \citep[e.g.][]{Bahe2015,Pallero2019,Donnari2021,Stevens2021}. The top panels show that the slow decline persisted for about few Gyr post infall, followed by a rapid drop in SFR, demonstrating that these galaxies underwent `slow-then-rapid' quenching.

Based on similar curves for different mass bins, we find that the average active time, $t_{\rm act}$\footnote{This is similar to the quenching timescale in exponential quenching models \citep{Wetzel2013,Hahn2017,Foltz2018}, but is different from the quenching timescale based on a threshold sSFR \citep[e.g.][]{Furlong2015,Oman2016,Wright2019,Akins2022}, or transition between two populations in colour \citep[e.g.][]{Trayford2016,Nelson2018,Wright2019}.} -- i.e. the duration between $t=t_{\rm c}$ and the moment of sharpest drop in the SFR (apparent in Fig.~\ref{xvstsi}) -- is not constant for all stellar masses and redshifts. At a fixed redshift, the median $t_{\rm act}$ increases monotonically with mass until $M_\star\approx 10^{9.9}\,{\rm M}_\odot$ and remains nearly constant for higher masses. The constant is lower for higher redshifts,
ranging from the median $t_{\rm act}\approx 3$~Gyr for $z=0$ to $t_{\rm act}\approx 1.35$~Gyr for $z=0.7$, deviating by $\approx 0.25$~Gyr between masses. This is, in fact, consistent with the evolution of $t_{\rm cross}$ which reduces from $\approx 2.9$~Gyr at $z=0$ to $\approx 1.6$~Gyr at $z=0.7$, in agreement with observations \citep{Mok2014,Fossati2017,Lemaux2019,Rhee2020}. We would like to point out, however, that `slow-then-rapid' quenching is not an accurate description of the evolutionary histories of galaxies with $M_\star \lesssim 10^{9.5}\,{\rm M}_\odot$, where
the `slow' phase is practically absent with $t_{\rm act}\lesssim 1$~Gyr.

These results favour a quenching mechanism that is largely independent of redshift and mass. Since it takes some time after infall for a galaxy to reach the dense regions of a cluster, it is expected to be influenced more by starvation/strangulation 
until the first pericentric passage, and then undergo ram-pressure and tidal stripping. Overall, this is expected to quench the galaxy over a duration similar to the crossing timescale, which is evident in the variation of $t_{\rm act}$ with redshift for $M_\star\gtrsim 10^{9.9}\,{\rm M}_\odot$. Apart from indicating mass-invariance, the constant $t_{\rm act}$ in this regime for a fixed redshift rules out AGN feedback as the primary quenching process for cluster satellites, especially at $M_\star\gtrsim 10^{10.5}\,{\rm M}_\odot$, where one might expect it to be relevant \citep{Mitchell2020}. At low masses (like $M_\star \lesssim 10^{9.5}\,{\rm M}_\odot$), a mass-dependence emerges likely because gravitational potentials become considerably shallower,
which is conducive for efficient ram-pressure stripping soon after infall, and expected to quench within $\approx 1$~Gyr \citep{Roediger2005,Steinhauser2016}.

Note that the SF subsample has not evolved like a typical cluster galaxy. As shown by the solid orange curves in Fig.~\ref{xvstsi}, their SFRs did not decline until infall, and the rate of decline was always shallower than the average cluster galaxy. Furthermore, we find some SF cluster galaxies that have not quenched even after being bound to clusters for durations longer than the typical active time. By selection, SF galaxies have been more robust against external quenching mechanisms than any typical galaxy that is exposed to cluster environment. We examine the group membership of our cluster galaxies across time, and find that most of them were centrals $\gtrsim 2$~Gyr before infall, whether we consider the full cluster sample or the SF one. However,
while the SF systems remained centrals until infall into a cluster, an average cluster galaxy was a satellite in a lower mass group for about 2 Gyr before being accreted by a cluster. This means that, unlike the SF subsample, most cluster galaxies have been pre-processed in such groups, which explains the earlier commencement of their molecular gas loss and SFR's decline.

In addition, the dashed curves in Fig.~\ref{xvstsi} show that the $M_\Hmol$'s evolution always traces that of SFR, regardless of stellar mass or redshift, which means that the impact of molecular gas loss on the SFR is nearly instantaneous. Another way to interpret this is that the cluster environment has negligible impact on the star formation efficiency of galaxies,
at least for the temporal resolution of the simulation. We find this to hold true across stellar masses.

\section{Uncovering the physical origin of the decline in the $\Hmol$ content of SF cluster galaxies}\label{trackparts}
In this section, we investigate how various processes influenced the $\Hmol$ content of SF cluster galaxies before and after infall. Through this analysis, we aim to highlight the mechanisms that have a stronger bearing on the evolution of $\Hmol$ in
such galaxies.

\subsection{Contributions from phenomenological modes}
In \eagleNS, when a gas particle gets converted into a star particle, \textit{all} of its mass gets transferred to the star particle, not just the mass that is contained in $\Hmol$ (which is post-processed). This means that, even if star formation is the only process affecting the ISM, a reduction in $M_\Hmol$ is \textit{not} equal to the increase in $M_\star$. Fortunately, since every particle has a unique ID in \eagleNS, it is possible to identify the gas particles that were converted into stellar particles by comparing their IDs between successive snapshots. Likewise, we can also identify the gas particles that were removed from the galaxy, those that were accreted by the galaxy, and the ones that remained in the galaxy but incurred a change in their $\Hmol$ mass due to, for example, variation in the local ionisation field, mass loss from stars, or radiative cooling/heating. Thus, 
particle-tracking is a viable technique for understanding how distinct modes govern gas flows 
to-, from- and within galaxies in \eagleNS.

\begin{figure*}
\begin{center}
  \includegraphics[width=1.95\columnwidth,trim={1cm 0.1cm 0.1cm 0.1cm}]{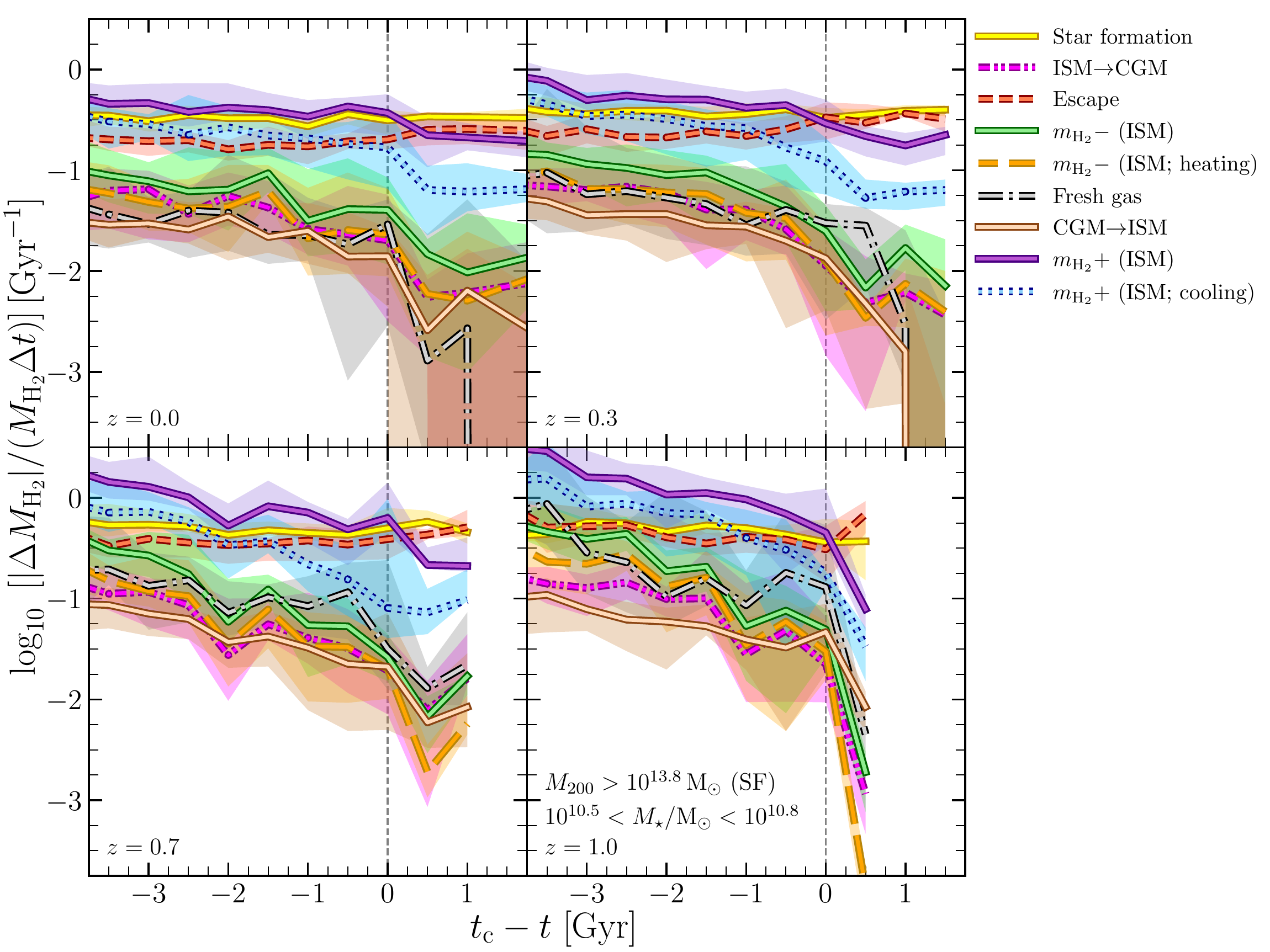}
   \caption{This shows how various processes influenced the $\Hmol$ content of cluster galaxies with 
   stellar masses in the range $10^{10.5}<M_\star/{\rm M}_\odot<10^{10.8}$ throughout their history. The solid yellow curve
   is the median specific-loss-rate of $\Hmol$ due to star formation. The dashed-double-dot magenta curve
   is a similar rate caused by movement of gas particles from the ISM to the CGM. The dashed red curve is the loss rate due to gas particles escaping the galaxy's subhalo (getting unbound). The solid
   green curve shows the loss rate due to reduction in the $\Hmol$ mass ($m_\Hmol$) of particles in the
   ISM. The dashed orange curve is the contribution of particle-heating to the solid green curve. 
   The dashed-dot grey curve is the replenishment rate due to accretion of fresh gas by the subhalo. 
   The solid cream curve shows the replenishment rate due to increase in $m_\Hmol$ as the CGM
   particles get incorporated into the ISM. The solid purple curve is the replenishment rate
   corresponding to the increase in $m_\Hmol$ of the particles that remain in the ISM. The dotted
   blue curve is the contribution of cooling to the solid purple curve. For each of these
   curves, the shaded region with the same colour encloses 20th to 80th percentiles. The $x$-axis is the same as in Fig.~\ref{xvstsi}. Before entering a cluster, the gain rate due
   to increase in $m_\Hmol$ was higher
   than the loss rates, but the there was a drop in the former a few Gyr prior to infall. At a moment close to the time of infall, star formation and escape began to dominate, which led to the steep decline in the $\Hmol$ content post infall.}
   \label{dmh2}
\end{center}
\end{figure*}

Note that the mass added via gas accretion cannot be estimated by simply identifying the gas particles that did not exist in the subhalo in the previous snapshot, because this does not account for changes incurred due to transport of gas from the circumgalactic medium (CGM) to the ISM. Thus, we need to distinguish between the two
for the analysis that follows. Since there is no physical boundary that separates them, this can be carried out in a variety of ways. The interstellar gas is generally denser, cooler, star-forming, and located closer to the halo centre. Consequently, methodologies adopted in the literature usually involve enforcing a cut on star formation rate \citep[e.g.][]{Wijers2020} and/or radial distance \citep[e.g.][]{van2017,DeFelippis2020,Fielding2020} coupled with that on density \citep[e.g.][]{Alcazar2017,Correa2018,Hafen2019,Mitchell2020}, temperature \citep[e.g.][]{Correa2018,Mitchell2020,Wright2022} or rotational support \citep[e.g.][]{Mitchell2018,Smithson2021}. We follow the approach delineated in \citet{Wright2022}, where we first determine the radial extent of the galaxy, $R_{\rm bary}$: the smallest radius\footnote{This is computed using
the {\tt BaryMP} module at \url{https://github.com/arhstevens/Dirty-AstroPy/blob/master/galprops/galcalc.py}.} at which the cumulative baryonic-mass profile based on the stars and all the gas that is cool ($T<10^{4.3}\,{\rm K}$) or star-forming (${\rm SFR>0}$) in the subhalo resembles an isothermal gas halo (i.e. the density follows $\rho_{\rm bary}\propto r^{-2}$; see \citealt{Stevens2014}). The ISM is then defined as the cool or star-forming gas that lies within $R_{\rm bary}$, and the rest of the subhalo gas is the CGM. Our visual inspection of the resultant gas distributions suggests that this method is indeed effective.

Once we have categorised the particles into the two classes, we track the stellar, gas and BH particles of the main progenitor and, for each pair of successive snapshots, 
determine the amount of $\Hmol$ that is
\begin{itemize}
    \item a) lost due to star formation,
    \item b) lost when particles move from the ISM to the CGM,
    \item c) lost due to particles escaping the system (or getting unbound),\footnote{We do not segregate these further
    based on escape caused by feedback and stripping, as doing so on a per-particle basis is non-trivial. Using a temperature threshold (e.g. $T>10^{7.5}\,{\rm K}$) to identify the gas particles that were heated by a feedback event will not capture the heating caused by events that occur at instances between two snapshots. 
    Moreover, even if we were to identify such particles, this does not guarantee that they were solely affected by feedback because both stripping and feedback can, in principle, impact a particle simultaneously. It is possible to estimate the efficacy of ram pressure by comparing it to the gravitational restoring-force exerted by the galaxy \citep[e.g.][]{Marasco2016,Manuwal2022}, but this can only be done reliably for a region rather than an individual particle.}
    \item d) lost due to a decrease in the $\Hmol$ content of particles in the ISM,
    \item e) added due to introduction of fresh gas in the system,
    \item f) added when particles in the CGM get accreted by the galaxy, and
    \item g) added due to increase in the $\Hmol$ content of particles in the ISM.
\end{itemize}

\noindent For a), we identify the progenitor's gas particles that exist as star particles in the descendant, and determine their cumulative $\Hmol$ mass. For b), we select the gas particles in the progenitor's ISM that are in the descendant's CGM, identify the ones whose $\Hmol$
mass  ($m_{\rm H_2}$) had reduced, and add up their deficits. For c), we identify the progenitor's gas particles that are not present in the descendant as either gas, stellar or BH particles, and compute their cumulative $\Hmol$ mass. For d), we select the gas particles that are common in the ISMs of the progenitor and the descendant, identify the ones that lost their $\Hmol$, and calculate the cumulative deficit. For e), we identify the descendant's gas particles that are not present in the progenitor, and add
up their $\Hmol$ mass. For f), we identify the CGM 
particles in the progenitor that are present in the descendant's ISM, select the ones whose $m_{\rm H_2}$
had increased, and compute their cumulative mass-credit. Lastly, for g), we select the ISM particles
that are common between the progenitor and the descendant, identify those that had an increase in
$m_{\rm H_2}$, and add up their credits.

One may notice that we have not alluded to the role of BHs above. Given that BHs accrete material that lies in their proximity, some $\Hmol$ can be lost to this 
mechanism. Likewise, some mass may be lost due to conversion of gas particles to BHs. It is not possible to identify individual gas particles that were accreted by a BH in \eagleNS,
but the data does provide the cumulative accreted-mass of each BH particle -- meaning one can calculate the \textit{total} gas mass that was lost via BH accretion. We find this to be $\sim 10^6\,{\rm M}_\odot$ at max, implying that the $\Hmol$ component of the lost gas is negligible. Comparisons of particle IDs reveal that the $\Hmol$ mass lost through conversion of gas particles to BHs is also negligible. Additionally, it is possible that some new gas particles were captured by the subhalo and then converted to stars between successive snapshots. Although these would not be identified in our analysis, they do not cause any \textit{net} change in the $\Hmol$ content and can therefore be ignored.

For each mode from a) to g), Fig.~\ref{dmh2} shows the median $\Hmol$ gain/loss rate divided by the $\Hmol$ mass (or the specific gain/loss rate) against the time since infall for the SF cluster galaxies with $10^{10.5}<M_\star/{\rm M}_\odot<10^{10.8}$, where each panel corresponds to a specific sample-redshift (as in previous figures). We have plotted the specific rates in particular because these are especially suited for assessing the efficacy of a process in
governing the gas content \citep[e.g. see][]{Wright2022}. The shaded regions encompass 20th to 80th
percentiles. The vertical dashed line corresponds to the moment when the look-back time $t=t_{\rm c}$. 

The results indicate that, before a typical galaxy entered a cluster (i.e. leftwards of the vertical line), $\Hmol$ was mainly replenished via increase in $m_\Hmol$ of the ISM particles (solid purple), and star formation was the main mode of loss (solid yellow), for all redshifts. The two modes of gas accretion (solid cream and dashed-dot grey) caused replenishment rates that were lower
than the contribution from ISM particles gaining $\Hmol$, by $\gtrsim 0.75$ dex, and therefore had a relatively weaker impact on the $\Hmol$ content. Similarly, the loss rate corresponding to the movement of gas from ISM to CGM (dashed-double-dot magenta) was about three times lower than that due to gas escape (dashed red). 

As the galaxies approached infall, the replenishment rate due to particles gaining $\Hmol$ 
had decreased, whereas the star-formation-driven depletion rate remained constant and approached the former. The latter is particularly interesting, because it suggests that the reduction in SFR was caused by lowering of the $\Hmol$ content rather than more efficient star formation, echoing the result in Fig.~\ref{xvstsi}. Meanwhile, the escape-driven rate increased by $\approx 0.25$ dex relative to 
its value a Gyr prior to $t_{\rm c}$, and thereby, became similar to the star-formation-driven rate.

Post infall (rightwards of the vertical line), the losses caused by star formation and escape dominated over the replenishment modes, and there was a sharp decline in the replenishment of $\Hmol$ via gas accretion, as is expected for galaxies in cluster environments \citep{van2017,Wright2020,Wright2022}. The
curve corresponding to the introduction of fresh gas (dashed-dot grey) truncates $\approx 1$ Gyr after infall, because gas supply is cut off for most galaxies. Overall, this shows that the steeper decline of $M_\Hmol$ of cluster galaxies post infall is a result of the dominance of star formation and gas escape over all the modes that replenish $\Hmol$.

There are some additional insights that can be derived from this analysis. For instance, it is worth noting that not all the $\Hmol$ gained as a result of increase in $m_\Hmol$ 
of ISM particles can be owed to cooling. In fact, as shown using the dotted blue curve, cooling accounts $\lesssim 50$ per cent at any given moment. Heating (dashed orange) has similar contribution to the decrease in $m_\Hmol$ of ISM particles. On the other hand, we find that all the transport from CGM to ISM is accompanied by cooling, the movement of gas from ISM to CGM always involves heating, and almost all the escaped $\Hmol$ comes from ISM. Similar results are obtained for other mass bins.

\subsection{Significance of stellar/AGN feedback (or a lack thereof)}\label{feedsig}
While it is clear that gas escape was one of the major modes of $\Hmol$ loss post infall, whether
this was primarily caused by stellar feedback, AGN feedback, or direct stripping of $\Hmol$, remains abstruse. In this regard, we assess the impact of the feedback mechanisms by comparing the injected energy against the binding energy of gas particles. 

In order to calculate the cumulative energy released due to stellar feedback between successive snapshots, we first identify all the gas particles that were converted to stellar particles, and note the mass of each such stellar particle at the time of its formation ($m_{\star}$). Then, the injected energy (in ergs) is determined as \citep{Schaye2015},
\begin{equation}
E_{\star}=8.73\times 10^{15}\sum_{i=1}^N{f_{{\rm th},i} m_{\star,i}},   
\end{equation}
where $f_{\rm th}$ is taken from the \eagle database and $N$ is the number of particles. Likewise, the energy injected
by the black hole is computed as
\begin{equation}
E_{\rm agn}=0.015c^2 m_{\rm acc},  
\end{equation}
where $c$ is the speed of light in vacuum and $m_{\rm acc}$ is the total mass accreted by the black hole between the successive snapshots.

We quantify the total binding energy of gas particles prior to their escape. Since the aim is to examine the impact of feedback on the $\Hmol$ loss caused by gas escape, we focus on the particles that were lost through this mode. Amongst these, we exclude the particles that did escape but were not representative of $\Hmol$. To this end, we rank order the escaped particles in decreasing order of their $\Hmol$ mass in the first snapshot, and determine the cumulative mass profile by summing up from lowest to highest rank. Then, we select all the particles with ranks below the rank corresponding to 90 per cent of their total $\Hmol$ mass, and add up their binding energies to obtain $E_{\rm bind}$.

On examining the evolution of $E_{\star}/E_{\rm bind}$ (not shown) for the galaxies pertaining to Fig.~\ref{dmh2}, we find that the $E_{\star}/E_{\rm bind}$ remained fairly constant at $\approx 10^{-13}$ throughout the history of our galaxies. The $E_{\rm agn}/E_{\rm bind}$, on the other hand,
had an erratic history where it was either $0$ or $\lesssim 10^{-3}$. Therefore, neither of the feedback mechanisms were efficient enough to eject the molecular gas by themselves, before, or after infall into clusters. This also holds true for other stellar
masses, including $M_\star\lesssim 10^{9.5}$ where quenching is generally fast. Thus, it appears that the energy required for escape of $\Hmol$ from SF cluster galaxies was essentially provided by stripping mechanisms that they were subjected to after infall into clusters.

\section{Summary}\label{summary}
Since the detection of $\Hmol$ requires observed galaxies to possess sufficient quantities of cold gas, most of the detected systems tend to be SF galaxies. The observational data for
such galaxies in clusters, particularly at high redshifts, remains sparse, precluding any overall consensus regarding the relationship between the $\Hmol$ content of SF galaxies, the cluster environment, and redshift. In this paper, we investigated this relationship by leveraging the 100 Mpc ``REFERENCE'' run from the \eagle project \citep{Schaye2015,Crain2015}. We selected SF galaxies at four different redshifts ranging from $z=0$ to $z=1$, and obtained their $\Hmol$ content in post-processing. We divided each of these samples into cluster and non-cluster
galaxies based on the mass of their groups, and examined differences in their $\Hmol$ content. We then investigated potential factors that modulate these differences, and their implications for galaxy evolution in clusters. Finally, we used particle tracking to reveal how the evolution of the $\Hmol$ content of SF cluster galaxies relates to star formation, molecular gas stripping, gas accretion, stellar/BH feedback, and other miscellaneous processes within the ISM.

The main results from this study are summarised as follows:
\begin{enumerate}
    \item At a fixed stellar mass, SF galaxies in clusters generally possess less $\Hmol$ than their non-cluster counterparts,
    but more $\Hmol$ than a typical cluster galaxy (Fig.~\ref{mh2vsmstar_allz}). This provides theoretical support for observational studies that claim the same. We found that the offset in $\Hmol$ masses between the two samples increases at lower redshifts at a given stellar mass, and also varies with stellar mass at a given redshift: it is highest for $M_\star\approx 10^{10.5}\,{\rm M}_\odot$ (upto $\approx 0.5$~dex), but diminishes below $M_\star\approx 10^{9.3}\,{\rm M}_\odot$.
    
    \item The $M_\Hmol$ histories of galaxies based on their main progenitors show that SF galaxies
    outside clusters did not possess more $\Hmol$ to begin with, and the $\Hmol$-deficit in SF cluster
    galaxies is a consequence of the steep decline of their $M_\Hmol$ relative to non-cluster counterparts around their time of first infall into a cluster (Fig.~\ref{mh2vst}). This decline, and the infall, commenced earlier for most cluster galaxies. The offset between $\Hmol$ content of the two samples (shown in Fig.~\ref{mh2vsmstar_allz}) is positively correlated to the time elapsed since infall for the cluster galaxies (see Fig.~\ref{dmh2vstsinf}), though there are other factors that modulate the scatter in this trend -- e.g. group mass,
    density at the virial radius, crossing timescale (Table~\ref{pca}, Section~\ref{modfact}).
    
    \item The evolution of $\Hmol$ content and SFR of cluster galaxies with respect to their first infall into a cluster showed that the two followed each other (Fig.~\ref{xvstsi}), implying a fairly invariant star-formation efficiency
    before and after infall. We found that an average cluster galaxy in \eagle with $M_\star \gtrsim 10^{9.5}\,{\rm M}_\odot$ undergoes a `slow-then-rapid' quenching, whereas those at lower masses quench rapidly. Most cluster galaxies were pre-processed in groups for $\approx 2$~Gyr before entering a cluster, but the SF ones, being centrals, were not. This 
    enabled better preservation of $\Hmol$ in SF cluster galaxies. 
    
    \item By tracking the gas and stellar particles in the main progenitors of SF cluster galaxies, we were able to calculate the changes in $\Hmol$
    mass that were caused by star formation, gas escape, gas accretion, and variations in the $\Hmol$ mass of gas particles in the ISM. We found that, before entering a cluster environment, the $\Hmol$ mass of an average SF galaxy was lost mostly to star formation and gas escape,
    but was compensated to a large extent by other processes; mainly by increase in $m_{\rm H_2}$
    of ISM particles (see Fig.~\ref{dmh2}). The sharp decline
    in the $\Hmol$ content since the time of infall occurred due to the domination of star formation and escape over the modes that replenished $\Hmol$. We compared the energy injected by feedback events against the binding energy of the escaped gas to infer that they were primarily removed via environmental stripping (Section~\ref{feedsig}). 
    \end{enumerate}

We foresee improvements and exploration beyond the analysis presented in this paper. The limited resolution of \eagle precludes self-consistent treatment of all the different phases of the ISM, but modelling these phases is crucial for properly capturing the 
physics germane to the formation and destruction of molecular gas. Although the post-processing prescription used in this work does account for such processes \citep{Stevens2019}, a self-consistent approach with better mass resolution would be more accurate. The analysis can be repeated with other cosmological hydrodynamical simulations [e.g. Horizon-AGN \citep{Dubois2014}, IllustrisTNG \citep{Pillepich2018}, FIRE-2 \citep{Hopkins2018}, Simba \citep{Dave2019}], to investigate whether our conclusions are generic to galaxies that form in a $\Lambda$CDM cosmology, or if they are specific to \eagleNS. Furthermore, the number statistics of SF cluster galaxies can be improved upon by using a box size larger than 100 Mpc, which would expand the scope to redshifts greater than $z=1$ and more massive clusters, and also help make predictions for the high-redshift science that can be carried out with the James Webb Space Telescope (\textit{JWST}).

\section*{Acknowledgements}
We would like to thank Aaron Ludlow for useful discussions and insights. AM acknowledges support from the Australian Government through a Research Training Program (RTP) Scholarship. 
ARHS is the recipient of the Jim Buckee Fellowship at The University of Western Australia. We acknowledge the Virgo Consortium for making their simulation data available. The \eagle simulations were performed using the DiRAC-2 facility at Durham, managed by the ICC, and the PRACE facility Curie based in France at TGCC, CEA, Bruy\`{e}res-le-Ch\^{a}tel. The
data was processed on the OzSTAR supercomputer, which is managed through the Centre for Astrophysics \& Supercomputing at the Swinburne University of Technology, and funded by Astronomy Australia Limited
and the Australian Commonwealth Government. The analysis was performed with the help of \textsc{\large astropy} \citep{Astropy}, \textsc{\large h5py}, \textsc{\large matplotlib} \citep{Hunter2007}, \textsc{\large numpy} \citep{Numpy}, \textsc{\large scikit-learn} \citep{Pedregosa2011}, and \textsc{\large scipy} \citep{Scipy} packages for \textsc{\large python}. The paper has been typeset using Overleaf.\footnote{\url{https://www.overleaf.com/}}

\section*{Data Availability}
The \eagle data used in this work is publicly available at \url{http://icc.dur.ac.uk/Eagle/database.php}, and the
descriptions for halo catalogues and particle data are provided in \citet{Mcalpine2016} and \citet{eaglepart}, respectively. The $\Hmol$ masses used in this work can be obtained through a reasonable request to the corresponding author. The observational data is accessible through the links provided in the respective papers. 



\bibliographystyle{mnras}
\bibliography{citations} 




\appendix
\section{Obtaining the $\Hmolb$ content of gas particles}\label{postproc}

\subsection{Computing the neutral hydrogen content}\label{neuthyd}
To determine the neutral hydrogen content, we use the \citet{Rahmati2013} prescription which was calibrated using the \textsc{traphic} \citep{Pawlik2008} radiative transfer simulations, and accounts for photoionisation due to the extragalactic ultraviolet background (UVB), self-shielding, radiative recombination, and collisional ionisation. We begin by computing the total photoionisation rate $\Gamma_{\rm phot}$ using equation (A1) and Table A1 of \citet{Rahmati2013}. For this, the hydrogen number density 
($n_{\rm H}$) is determined using the hydrogen fraction (i.e. the fraction of gas mass in hydrogen, $X$) and the particle mass provided in the \eagle database. We assume the UVB as \citet{Haardt2012}, which is weaker by a factor of 3 compared to the UVB adopted for \eagle (i.e. \citealt{Haardt2001}), but
nonetheless results in similar masses \citep{Bahe2016,Crain2017}. In \eagleNS, the temperature of a star-forming gas particle is unrealistic and set by the polytropic equation of state mentioned in Section~\ref{sim}. For such particles, we adopt a temperature $T=10^4\,{\rm K}$ in order to mimic the warm, diffuse ISM around young stellar populations \citep[cf.][]{Crain2017}. Finally, we determine the fraction of hydrogen that is neutral ($f_{\rm n}$) by inserting the values for $\Gamma_{\rm phot}$, $T$ and $n_{\rm H}$ in equation (A8) of \citet{Rahmati2013}.

\subsection{Deriving the contribution from $\Hmol$ to the neutral hydrogen content}\label{molfrac}
Next, we partition the neutral hydrogen into atomic and molecular components. We inform our choice of partitioning scheme -- used for deriving the atomic and molecular contributions to gas particles in \eagle -- based on some comparative tests that were done on the five prescriptions mentioned in Section~\ref{sim}. In this section, we provide short descriptions for the prescriptions and their implementation.

\begin{itemize}
    \item BR06: This is a scaling relation between molecular-to-atomic gas ratio and the mid-plane gas pressure ($P$) that was derived empirically using observations of 14 galaxies. The relation is given as
    \begin{equation}
    R \equiv \frac{\Sigma_{\Hmol}}{\Sigma_{\HIs}} = \left(\frac{P}{P_0}\right)^\alpha\,,    
    \label{br06}
    \end{equation}
    where $\Sigma_x$ is the surface density of component $x$, $P_0/k_{\rm B}=4.3\times10^4~\cc~\K$ and $\alpha=0.92$. We estimate the gas pressure in each gas particle based on its density, $n_{\rm H}$, and temperature, $T$; i.e. $P/k_{\rm B}=n_{\rm H}T$, where $k_{\rm B}$ is the Boltzmann constant.
    \item L08: Similar to BR06, this is also an empirically derived relation between $R$ and $P$ but based on a relatively larger sample of 23 galaxies. It follows the same form as equation~(\ref{br06}) but differs in the value of constants, which are slightly lower at $P_0/k_{\rm B}=1.7\times10^4~\cc~\K$ and $\alpha=0.8$.
    \item GK11: This prescription is based on various ``fixed ISM'' hydrodynamical simulations that were run using adaptive mesh refinement (AMR), each with a fixed dust-to-gas ratio (same value for all) and interstellar radiation field at $1000$~\AA. The simulations included gas dynamics, radiative transfer of UV photons from stellar particles, a chemical network of hydrogen and helium with non-equilibrium cooling and heating rates, and molecular hydrogen formation on dust grains. 
    
    There are two sets of fitting formulae in their
    paper that can be used to derive $f_{\Hmol}$. The first is their equation~(6):
    \begin{equation}
    f_{\Hmol} = \frac{1}{1+\exp(-4x-3x^3)}\,,
    \label{gkeq1}
    \end{equation}
    where
    \begin{equation}
    x \equiv \Lambda^{3/7}\ln\left(D_{\rm MW} \frac{n_{\rm H}}{\Lambda n_*}\right)\,.
    \end{equation}
    $n_*=25~\cc$, $D_{\rm MW}$ is the dust-to-gas ratio relative to the Milky Way value, and $\Lambda$ is given by
    \begin{equation}
    \Lambda \equiv \ln\left[1+gD_{\rm MW}^{3/7}\left(\frac{U_{\rm MW}}{15}\right)^{4/7}\right]\,. 
    \end{equation}
    Here $U_{\rm MW}$ is the normalisation of the interstellar far-UV (FUV) flux at $1000$~\AA, $g$ is a fitting formula used to account for the transition between the self-shielding regime ($g\approx 1$) to dust shielding regime ($g\propto D_{\rm MW}^{-1}$):
    \begin{equation}
    g \equiv \frac{1+\alpha s+s^2}{1+s}\,,    
    \end{equation}
    where
    \begin{equation}
    \alpha \equiv \frac{2.5 U_{\rm MW}}{1+(0.5U_{\rm MW})^2}\,,
    \end{equation}
    \begin{equation}
    s \equiv \frac{0.04}{D_\star+D_{\rm MW}}\,.    
    \end{equation}
    $D_\star$ denotes the transition point where the formation of $\Hmol$ mainly occurs through gas-phase reactions:
    \begin{equation}
    D_\star \equiv 0.0015\ln[1+(3U_{\rm MW})^{1.7}].    
    \end{equation}
    
    The second approach is given by their equation~(10):
    \begin{equation}
    f_{\Hmol} = \left(1+\frac{\Sigma_{c}}{\Sigma_{\HIs+\Hmol}}\right)^{-2}
    \label{gkeq2}
    \end{equation}
    where $\Sigma_{\rm c}$ is the characteristic neutral-gas surface density where the relationship between SFR surface density and molecular hydrogen surface density steepens. It is given by equation~(14) in their paper, i.e.
    \begin{equation}
    \Sigma_{\rm c} = \frac{20\Lambda^{4/7}}{D_{\rm MW}\sqrt{1+U_{\rm MW}D^2_{\rm MW}}}.    
    \end{equation}
    \item K13: This is an extension of the theoretical model developed by \citet{Krumholz2008atomic,Krumholz2009atomic,Krumholz2009} for accurate treatment of $\HI$ dominated regions. Under this model, a galactic gas disk is partitioned between $\HI$ (non star-forming) and $\Hmol$ (star-forming) based on the gas column density, metallicity and the ratio of the interstellar radiation field (ISRF) to the cold atomic ISM density. If the pressure implies co-existence of warm and cold atomic phases, the ratio of ISRF to density is nearly constant, and the conversion is only a function of column density and metallicity. For other regions with low ISRF and long depletion timescales, the ratio is not fixed and there is a density floor imposed in order to maintain hydrostatic equilibrium. 
    
    Similar to \citet{Smithson2021}, we apply equation~(10) of \citet{K13} to determine molecular gas fractions of our gas particles:

\begin{equation}
f_{\Hmol} = \left\{
\begin{array}{lr}
1 - 3\, S\, (4 + S)^{-1} & \forall~S<2\\
0 & \forall~S \geq 2
\end{array}
\right.
\,,
\end{equation}
where
\begin{equation}
S \equiv \frac{\ln( 1 + 0.6\, \chi + 0.01\, \chi^2)}{0.6\, \tau_{\rm c}}\,,
\end{equation}
\begin{equation}
\tau_{\rm c} \equiv 0.066\, f_{\rm c}\, D_{\rm MW}\, \frac{\Sigma_{\rm H\,{\LARGE{\textsc i}}+H_2}}{{\rm M}_{\odot}\,{\rm pc}^{-2}}\,,
\end{equation}
\begin{equation}
\chi \equiv 72\,U_{\rm MW} \left(\frac{n_{\rm CNM}}{{\rm cm}^{-3}}\right)^{-1}.
\end{equation}
Here $f_{\rm c}$ is a clumping factor to account for the difference between the physical scale of an atomic--molecular complex, and the scale at which it is observed, i.e. spatial resolution of the simulation (for \eagleNS, $f_{\rm c}=5$). $n_{\rm CNM}$ is cold neutral medium (CNM) density given by
\begin{equation}
n_{\rm CNM} \equiv {\rm max}[n_{\rm CNM,2p},\, n_{\rm CNM,hydro}]\,,
\end{equation}
where $n_{\rm CNM,2p}$ is the CNM density for a two-phase equilibrium between cold and warm neutral medium, i.e.
\begin{equation}
n_{\rm CNM,2p} \equiv 23\, U_{\rm MW}\, \frac{4.1}{1 + 3.1\, D_{\rm MW}^{0.365}}\, {\rm cm}^{-3}\,,
\end{equation}
and $n_{\rm CNM,hydro}$ is the lowest possible density for a CNM in hydrostatic equilibrium corresponding to the maximum temperature of $T_{\rm CNM,max}=243~\K$ \citep{Wolfire2003}; i.e.  
\begin{equation}
n_{\rm CNM,hydro} \equiv \frac{P_{\rm th}}{1.1\, k_B\, T_{\rm CNM,max}}\,.
\end{equation}
Here the factor of 1.1 accounts for He, and $P_{\rm th}$ is the thermal pressure. The mid-plane pressure ($P_{\rm mp}$) can be expressed as a sum of three contributions from: 1) the self-gravity of the $\HI$ that is not gravitationally bound, 2) the weight of the $\HI$ within the gravitational potential from the bound $\Hmol$ clouds, and 3) the weight of the $\HI$ within the gravitational potential provided by stars and dark matter. $P_{\rm th}$ can be written as a function of the sound speed in the WNM ($c_w$). $P_{\rm th}$ is smaller than $P_{\rm mp}$ by a factor of $\alpha\approx 5$ due to additional support from magnetic fields, cosmic rays and turbulence. Based on the expressions in \citet{K13}, this implies
\begin{equation}
P_{\rm th} = \frac{\pi\, G\, \Sigma_{\rm H\,{\LARGE{\textsc i}}}^2}{4\, A} \left( 1 + 2\,R + \sqrt{(1 + 2\,R)^2+\mathcal{F}} \right)\,,
\end{equation}
where
\begin{equation}
\mathcal{F} \equiv \frac{32\, \zeta_d\, A\, f_w\, c_w^2\, \rho_{\rm sd}}{\pi\, G\, \Sigma_{\rm H\,{\LARGE{\textsc i}}}^2}\,,
\end{equation}
and
\begin{equation}
R \equiv \frac{f_{\rm H_2}}{1 - f_{\rm H_2}}\,.
\end{equation}
$\rho_{\rm sd}$ is the density of DM and stars within the gas disk, and $\zeta_d\approx 0.33$ is a numerical factor that depends on the shape of the gas surface density profile.

\item GD14: \citet{GD14} provides two equations for deriving the $\Hmol$ content [equations (6) and (8)], but only equation (8) is appropriate for the spatial resolution of \eagle ($\approx 0.7$ kpc). It gives
\begin{equation}
R = \left( \frac{\Sigma_{\rm H\,{\LARGE{\textsc i}}+H_2}}{\Sigma_{R=1}} \right)^{\alpha},
\label{mainpres}
\end{equation}
where ${\Sigma_{\rm H\,{\LARGE{\textsc i}}+H_2}}$ is the neutral-hydrogen surface density, and $\Sigma_{R=1}$ is the surface density at which the atomic and molecular hydrogen fractions are equal:
\begin{equation}
\Sigma_{R=1} = \frac{50\, \sqrt{0.001 + 0.1\,U_{\rm MW}}}{g\, (1 + 1.69\, \sqrt{0.001 + 0.1\, U_{\rm MW}})}\, {\rm M}_{\odot}\, {\rm pc}^{-2}.
\end{equation}
Here $g$ is determined as
\begin{equation}
g \equiv \sqrt{D_\star^2 + D_{\rm MW}^2}\,,
\end{equation}
where $D_{\rm MW}$ is the dust-to-gas ratio relative to the Milky Way value and
\begin{equation}
D_\star \equiv 0.17\, \frac{2 + S^5}{1 + S^5}\,.
\label{dstar}
\end{equation}
$S$ is the sampling length of the simulation in units of 100 pc. $\alpha$ is calculated as
\begin{equation}
\alpha \equiv 0.5 + \left(1 + \sqrt{\frac{U_{\rm MW}\, D_{\rm MW}^2}{600}} \right)^{-1}.
\end{equation}
Thus, the value of $R$ is essentially set by the values of $\Sigma_{\rm H\,{\LARGE{\textsc i}}+H_2}$ and 
$S$. 
\end{itemize}

For implementing GK11, K13 and GD14, we follow the computational approach of \citet{Stevens2019}. The neutral-hydrogen surface density can be expressed as 
\begin{equation}
\Sigma_{\rm H\,{\LARGE{\textsc i}}+H_2} = f_{\rm n}X\Sigma,
\label{neutdens}
\end{equation}
where $\Sigma$ is the total-gas surface density. Since $\Sigma$ is a 2-dimensional quantity, whereas a gas particle in \eagle possesses a 3-dimensional density ($\rho$), we need to derive $\Sigma$ by multiplying $\rho$ with 
a measure of the particle's size. To this end, we assume the gas to be thermalised \citep{Schaye2008} and approximate its size as the Jeans length for an ideal gas, $L$ \citep{Schaye2001}. This can be written as
\begin{equation}
\Sigma = \rho L = \rho \sqrt{\frac{\gamma k_{\rm B}T}{\mu m_{\rm g} G}} = \sqrt{\frac{\gamma (\gamma-1)u\rho}{G}}\,,
\label{sig}
\end{equation}
where $\gamma$ is the ratio of specific heat capacity at constant pressure to that at constant volume, $k_{\rm B}$ is
the Boltzmann constant, $T$ is the particle temperature, $u$ is the specific internal energy, and $G$ is the gravitational 
constant. Assuming that all gas is thermalised, we approximate
\begin{align}
\gamma & = \frac{5}{3}(1-f_{\rm mol})+\frac{7}{5}f_{\rm mol}
\label{gamma}
\end{align}
and
\begin{align}
u & = [X+4(1-Z-X)]\left[\frac{1+(1-f_{\rm n})/f_{\rm n}}{(1-Z)(1+2(1-f_{\rm n})/f_{\rm n}-f_\Hmol/2)}\right]\,,
\label{u}
\end{align}
where $f_{\rm mol}$ is the ratio of $\Hmol$ to the total non-metallic component (H+He), i.e.
\begin{equation}
f_{\rm mol} = \frac{f_\Hmol f_{\rm n}X}{1-Z}.
\label{fmol}
\end{equation}
$f_\Hmol$ is the fraction of neutral hydrogen in molecular phase. 

\begin{figure*}
\begin{center}
  \includegraphics[width=2.2\columnwidth,trim={0.1cm 0.1cm 0.1cm 0.1cm}]{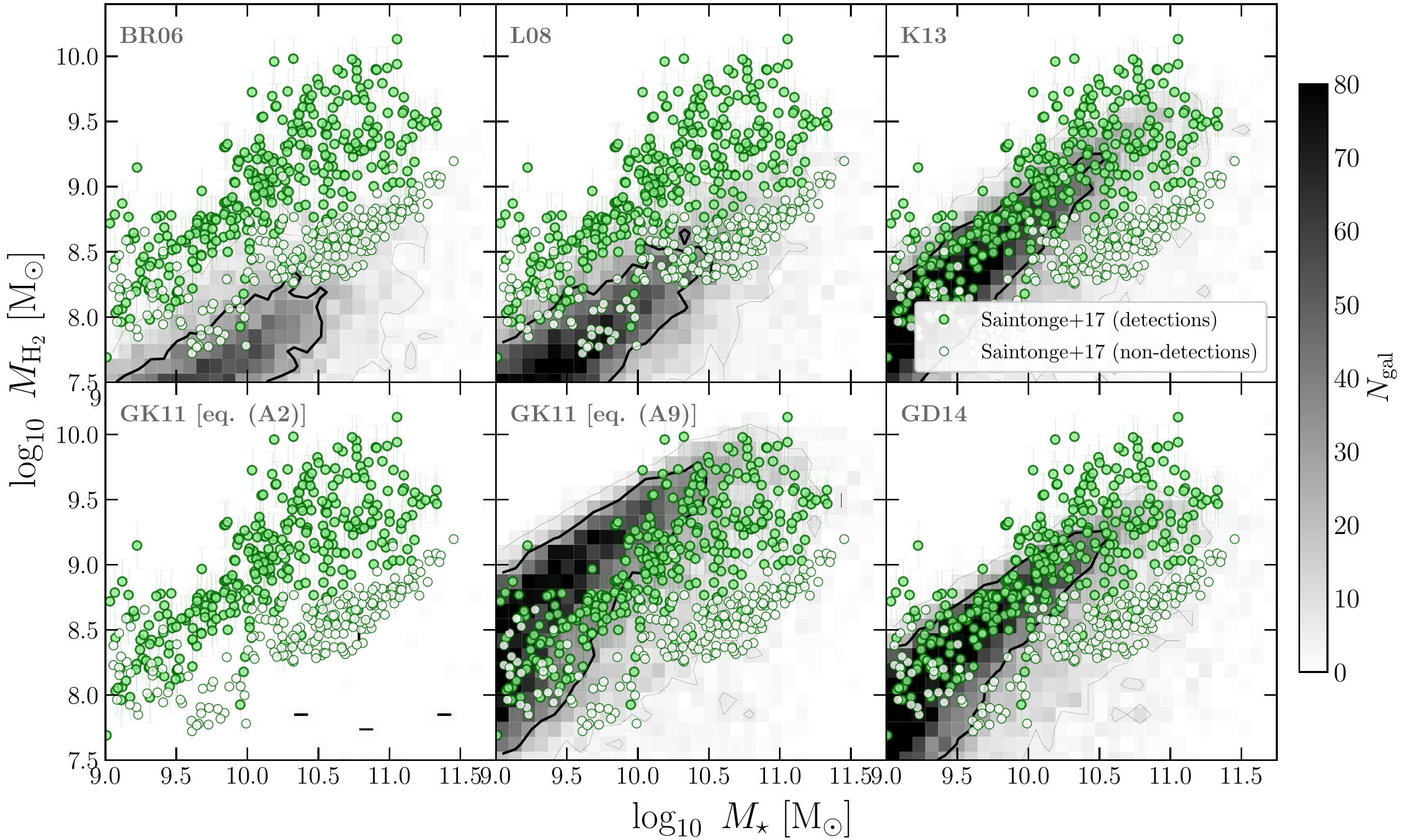}
   \caption{Molecular hydrogen masses plotted against stellar mass for galaxies with $M_\star>10^9\,{\rm M}_\odot$ in \eagle and 
   xCOLD GASS. In each panel, results from \eagle are shown as a 2-dimensional histogram where the number of galaxies in a bin ($N_{\rm gal}$) increases from white to black as shown in the colour bar. The black contours encompass 68 and 95 per cent of the \eagle data; the former is thicker. The $\HI/\Hmol$ partitioning scheme adopted is mentioned in the top-left corner. The two panels for the GK11 prescription correspond to equations~(6) and (10) in \citet{GK11} [equations~(\ref{gkeq1}) and (\ref{gkeq2}) in this paper, respectively]. The filled and open brown-circles are the $M_{\Hmol}$ values for detections and upper limits for non-detections in xCOLD GASS, respectively. The $\Hmol$ masses based on the L08 and BR06 prescriptions are significantly lower than those in the observations, while the K13 and GD14 produce $\Hmol$ masses that are broadly consistent with the observations.}
   \label{mh2xcold}
\end{center}
\end{figure*} 

As both $\gamma$ and $u$ are dependent on $f_\Hmol$, solving equation~(\ref{neutdens}) for $\Sigma_{\rm H\,{\LARGE{\textsc i}}+H_2}$ would need prior knowledge of $f_\Hmol$. We compute $f_\Hmol$ iteratively,
where it is initialised as $0$ for the first iteration and used to determine $f_{\rm mol}$ [equation~(\ref{fmol})]. This is used to get $u$ and $\gamma$ from equations~(\ref{u}) and (\ref{gamma}) which, along with equations~(\ref{sig}) and (\ref{neutdens}), give $\Sigma_{\rm H\,{\LARGE{\textsc i}}+H_2}$. Then, we obtain $f_\Hmol$ using the equations in the prescription, where we take the $\Sigma_{\rm H\,{\LARGE{\textsc i}}+H_2}$ just derived, $S=0.01L$, compute $U_{\rm MW}$ based on \citet{Diemer2018}, and assume $D_{\rm MW}$ to scale with metallicity and -- similar to \citet{Lagos2015} -- equal to the ratio of smoothed metallicity of the gas particle ($Z$) to the solar metallicity ($Z_\odot=0.0127$; \citealt{Allende2001}). This is repeated until $f_\Hmol$ converges within 0.5 per cent of its value in the previous iteration, or else, if there have been 300 iterations. 

\section{Identifying valid post-processing prescriptions}\label{prestest}
Now, we elaborate on the tests that were performed on these prescriptions. This includes a comparison against
molecular gas estimates for $z=0$ galaxies, and the cosmic $\Hmol$ density across redshifts.

\subsection{Comparison against observations at $z=0$}\label{obscomp1}
We compare the $\Hmol$ masses of $z=0$ galaxies predicted by the five models against the observed values from the xCOLD GASS survey \citep{Saintonge2017}, which is an extension of the COLD GASS survey \citep{Saintonge2011} 
to lower stellar masses and redshifts. It includes CO (1--0) observations of 532 galaxies at $0.01<z<0.05$. 
Amongst these, 366 galaxies (with $M_\star>10^{10}\,{\rm M}_\odot$ and $0.025<z<0.050$) are from COLD GASS, 
and the remaining 166 galaxies (with $10^{9}<M_\star/{\rm M}_\odot<10^{10}$ and $0.01<z<0.02$) were randomly 
selected from the SDSS group sample. This dataset serves as an ideal $z=0$ benchmark for galaxy evolution 
studies, as it is an unbiased and representative selection based purely on redshift and stellar mass, with 
an even sampling of the stellar mass space.

Fig.~\ref{mh2xcold} shows a scatter plot of $M_{\Hmol}$ against $M_\star$ for $z=0$ \eagle galaxies with stellar masses above $M_\star=10^9\,{\rm M}_\odot$, alongside galaxies in xCOLD GASS. The results from \eagle are shown as a 2-dimensional density plot, such that the density of galaxies increases from white to black. The black contours enclose 68 and 95 per cent of the \eagle data. The detections and upper limits from xCOLD GASS are shown as filled and open circles, respectively.

The prescriptions of BR06 and L08 predict considerably lower $\Hmol$ masses compared to observations. This discrepancy is potentially due to simplifying assumptions used by these models -- e.g. they do not account for relevant physical processes such as ionisation due to the local UV field, or the formation of molecular gas on dust grains. Also, the gas disk in the simulation has a larger scale height than expected \citep{Bahe2016,Llambay2018}, which causes the thermal pressure in high-density gas and $\Hmol$ content to be lower than a galaxy with a resolved cold phase \citep{Diemer2018}. This also explains, at least in part, the good agreement of $\HI$ masses from these schemes with observations reported by \citet{Manuwal2022}. Considering these factors, we choose to avoid these prescriptions.

We find that the two equations in GK11 [equations~(\ref{gkeq1}) and (\ref{gkeq2})] produce vastly different results, which indicates that this prescription may not be reliable for \eagleNS. The K13 and GD14 prescriptions produce $\Hmol$ masses that are in reasonable agreement with the data from xCOLD GASS, and are also consistent with each other. These prescriptions also produce sensible $\HI$ masses \citep{Manuwal2022}. This is reasonable, as these schemes are based on detailed analytical/numerical treatment of key physical processes, e.g. formation of $\Hmol$ on dust grains, and ionisation due to interstellar radiation field.

\subsection{Evolution of the cosmic $\Hmol$ density}\label{obscomp2}
Additionally, we estimate the co-moving cosmic $\Hmol$ density in \eagle for redshifts spanning $z = 0$ to $z = 4.5$\footnote{Note that we do not compute $\Hmol$ beyond this redshift because the \citet{Rahmati2013} prescription (used to obtain the neutral hydrogen) is only applicable until $z = 5$.} and compare them against the compilation of observed cosmic $\Hmol$ densities from ALMACAL \citep{Klitsch2019}, ASPECS LP \citep{Decarli2020,Magnelli2020}, \citet{Berta2013}, COLDz \citep{Riechers2019}, \citet{Garratt2021}, \citet{Maeda2017}, PHIBBS2 \citep{Lenkic2020}, \citet{Scoville2017}, VLASPECS \citep{Riechers2020} and xCOLD GASS \citep{Fletcher2021}.

\begin{figure}
  \includegraphics[width=0.98\columnwidth,trim={0.1cm 0.1cm 0.1cm 0.1cm}]{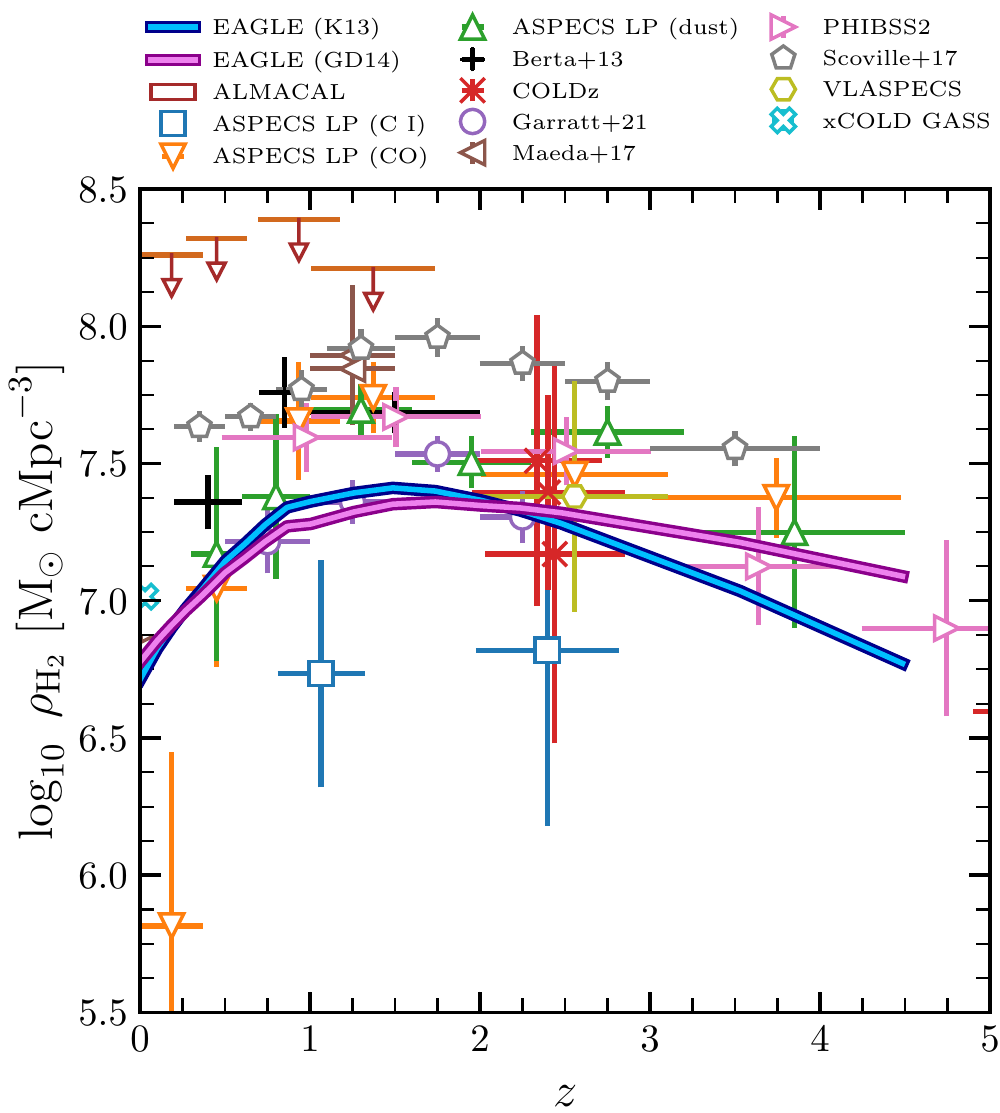}
   \caption{The co-moving cosmic $\Hmol$ density plotted as a function of redshift for \eagleNS, compared to various
   observational datasets. The blue and violet curves show \eagle results based on the K13 and GD14 prescriptions, respectively. 
   Estimates from various observations are shown as symbols with error-bars, and upper limits are shown as downward-pointing arrows. For the prescriptions considered here, 
   the evolution of the cosmic $\Hmol$ density is sufficiently consistent with the observed values for our purposes, considering the large observational uncertainties.}
   \label{rhoh2}
\end{figure}

Fig.~\ref{rhoh2} shows the $\Hmol$ density in \eagle as a function of redshift as predicted by 
the K13 and GD14 models (blue and violet curves, respectively), along with the observational values. Note that observed values carry large uncertainties and span a wide density range at any given redshift. If we take observations at face value, a comparison with the results
from \eagle suggests that both the K13 and GD14 prescriptions produce $\Hmol$ densities that are below the majority of observed values -- especially at $z=1$, where the offset is $\approx 0.4$ dex. We are unsure if this is primarily due to the limitations of the models or the observations, as observational estimates are likely to carry additional (systematic) uncertainties beyond the reported errors. For example, most of the surveys are biased towards gas-rich galaxies, especially at high redshifts. There is also uncertainty related to the CO-to-$\Hmol$ conversion factor ($\alpha_{\rm CO}$) used to estimate $\Hmol$ masses from CO flux. Nevertheless, we note that predicted $\Hmol$ densities from \eagle lie within the range spanned by the observed values at all redshifts. The shape of the curves is also consistent with the general trend in observations: an increase in the cosmic density until $z\approx 1.6$, and a decline at higher redshifts. We interpret this as an additional support for the reliability of these partitioning schemes.

\bsp	
\label{lastpage}
\end{document}